\documentclass[a4paper]{report}\usepackage[]{graphicx}\usepackage[]{color}
\makeatletter
\def\maxwidth{ %
  \ifdim\Gin@nat@width>\linewidth
    \linewidth
  \else
    \Gin@nat@width
  \fi
}
\makeatother

\definecolor{fgcolor}{rgb}{0.345, 0.345, 0.345}
\newcommand{\hlnum}[1]{\textcolor[rgb]{0.686,0.059,0.569}{#1}}%
\newcommand{\hlstr}[1]{\textcolor[rgb]{0.192,0.494,0.8}{#1}}%
\newcommand{\hlopt}[1]{\textcolor[rgb]{0,0,0}{#1}}%
\newcommand{\hlstd}[1]{\textcolor[rgb]{0.345,0.345,0.345}{#1}}%
\newcommand{\hlkwa}[1]{\textcolor[rgb]{0.161,0.373,0.58}{\textbf{#1}}}%
\newcommand{\hlkwb}[1]{\textcolor[rgb]{0.69,0.353,0.396}{#1}}%
\newcommand{\hlkwc}[1]{\textcolor[rgb]{0.333,0.667,0.333}{#1}}%
\newcommand{\hlkwd}[1]{\textcolor[rgb]{0.737,0.353,0.396}{\textbf{#1}}}%

\usepackage{framed}
\makeatletter
\newenvironment{kframe}{%
 \def\at@end@of@kframe{}%
 \ifinner\ifhmode%
  \def\at@end@of@kframe{\end{minipage}}%
  \begin{minipage}{\columnwidth}%
 \fi\fi%
 \def\FrameCommand##1{\hskip\@totalleftmargin \hskip-\fboxsep
 \colorbox{shadecolor}{##1}\hskip-\fboxsep
     \hskip-\linewidth \hskip-\@totalleftmargin \hskip\columnwidth}%
 \MakeFramed {\advance\hsize-\width
   \@totalleftmargin\z@ \linewidth\hsize
   \@setminipage}}%
 {\par\unskip\endMakeFramed%
 \at@end@of@kframe}
\makeatother

\definecolor{shadecolor}{rgb}{.97, .97, .97}
\definecolor{messagecolor}{rgb}{0, 0, 0}
\definecolor{warningcolor}{rgb}{1, 0, 1}
\definecolor{errorcolor}{rgb}{1, 0, 0}
\newenvironment{knitrout}{}{} 

\usepackage{alltt}
\usepackage[utf8]{inputenc}
\usepackage[T1]{fontenc}
\usepackage{RJournal}
\usepackage{amsmath,amssymb,array}
\usepackage{booktabs}

\usepackage{mathtools,mathpazo}

\usepackage{blkarray}

\usepackage{xspace}

\usepackage{placeins}

\usepackage{siunitx}

\renewenvironment{knitrout}{\vspace{0em}}{\vspace{-0.5em}}

\usepackage[disable]{todonotes} %

\newcount\Comments  
\usepackage{color}
\usepackage{adjustbox}
\newcommand{\kibitz}[2]{\ifnum\Comments=1\textcolor{#1}{#2}\fi}

\newcommand{\new}[1]  {#1}

\newcommand{\m}[1]{\ensuremath{#1}}
\newcommand{\n}[1]{_{\mathrm{#1}}}
\newcommand{\ns}[2]{_{\mathrm{#1},#2}}
\newcommand{\s}[2]{_{#1,#2}}

\newcommand{\mB}[1]{\mbox{\boldmath{$#1$}}}

\newcommand{\Ta}[1]{\m{T\ns{a}{#1}}} 
\newcommand{\mBTa}[1]{\m{\mB{T}\ns{a}{#1}}} 
\newcommand{\nk}{\m{n_k}\xspace} 

\let\proglang=\textsf

\newcommand{\rpackage}[2]{\proglang{\href{#2}{#1}}}

\newcommand{\onlineforecast}{\proglang{\href{https://onlineforecasting.org}{onlineforecast}}\xspace}
\newcommand{\Rprog}{\proglang{R}\xspace}
\newcommand{\Python}{\proglang{Python}\xspace}

\newcommand{\vignette}[1]{\href{https://onlineforecasting.org/vignettes/#1.html}{#1}}
\newcommand{\vignettecode}[1]{\href{https://onlineforecasting.org/vignettes/#1.R}{(code)}}
\IfFileExists{upquote.sty}{\usepackage{upquote}}{}
\begin{document}

\sectionhead{Contributed research article}
\volume{XX}
\volnumber{YY}
\year{20ZZ}
\month{AAAA}

\begin{article}
\title{onlineforecast: An R Package for Adaptive and Recursive Forecasting}

\author{by Peder Bacher, Hj\"orleifur G. Bergsteinsson, Linde Frölke, Mikkel L. Sørensen, Julian Lemos-Vinasco, Jon Liisberg, Jan Kloppenborg Møller, Henrik Aalborg Nielsen and Henrik Madsen}

\maketitle

\abstract{Systems that rely on forecasts to make decisions, e.g.\ control or
  energy trading systems, require frequent updates of the forecasts. Usually,
  the forecasts are updated whenever new observations become available, hence in
  an online setting. We present the \Rprog package \onlineforecast that provides a
  generalized setup of data and models for online forecasting. It has
  functionality for time-adaptive fitting of dynamical and non-linear
  models. The setup is tailored to enable the effective use of forecasts as model
  inputs, e.g.\ numerical weather forecast. Users can create new models for
  their particular applications and run models in an operational setting. The
  package also allows users to easily replace parts of the setup, e.g.\ using
  neural network methods for estimation. The package comes with comprehensive
  vignettes and examples of online forecasting applications in energy systems,
  but can easily be applied for online forecasting in all fields.}

\section[Introduction]{Introduction} \label{sec:intro}

\todo{see
  \url{https://awesomeopensource.com/projects/forecasting}}

Time series analysis and forecasting are of indispensable importance to numerous
practical fields such as business, finance, science and engineering \citep{cryer2008time}. Time series
analysis is the process of statistical modelling of time series, i.e.\ data
which is sampled at different points in time over a period -- often with a
constant distance in time, i.e.\ equidistant. Classical time series models for a
single equidistant time series use past values of the response variable (model
output) as predictors (inputs). In this way, appropriate models describing the
inherent auto-correlation structure of the time series can be realized. Such
models are exponential smoothing (e.g.\ Holt-Winters), AutoRegressive (AR) and
Moving Average (MA), and usually the combination of the latter two as ARMA
models. When multiple correlated time series are at hand, they can be used as
model inputs to improve forecasts. They are then called eXogenous variables and
the classical model becomes an ARMAX -- hence the \textit{X} indicates that
input variables are included.

The use of ARMAX models and variations thereof is widespread \citep{de200625},
especially in modelling of energy systems due to the high dependency between
e.g.\ weather, load, renewable generation and periodic phenomena. Load
forecasting is an obvious example. A nice overview for electric load forecasting
is given by \Citet{Hesham2002} and \citet{hong2016probabilistic}, and for heat
load by \Citet{Dotzauer2002277} who demonstrates the dependency between the
response variable, heat load, and the predictor, ambient temperature, using a
piecewise linear function. It is also proposed to model the daily and weekly
diurnal using hours of the week as inputs. For solar power forecasting
\citep{kleissl2013solar} the improvement from an autoregressive (AR) to an AR
with exogenous input (ARX), where the ARX model uses numerical weather
predictions (NWPs) as inputs, is demonstrated by \Citet{BACHER20091772}. The ARX
model uses past observations and NWPs of global irradiance to forecast the power
production from PV systems and the ARX model obtains higher accuracy than the AR
model. \Citet{bacher2013short} identifies exogenous variables that are suitable
for forecasting the heat load of a building, with similar models.

Energy systems are time-varying systems as they usually change over time due to
wear and contamination, like dirt on solar panels or changes in usage. For
example, with new tenants in a house the dependency between the heat load and
other variables, like calendar and temperature, changes. Therefore, a forecast
model needs to adapt -- the model coefficients are not optimal if they are
constant, they need to be updated and allowed to change over time. The Recursive
Least Square (RLS) method provides a recursive estimation scheme for the
coefficients in regression models, where they are updated at each step when new
data becomes available. Introduction of a forgetting factor in RLS
allows control on how fast the coefficients can change over time -- this is
referred to as adaptive recursive estimation, with exponential forgetting, in
linear regression and autoregressive models. The method is described by
\Citet{ljung1983theory} and the advances that has been made since then, see
e.g.\ \citep{engel2004kernel}.

\subsection{Time series modelling and forecasting in R}

A wide range of existing software useful for time series forecasting is
currently available -- all have their suitable applications
\citep{chatfield2019analysis,siebert2021systematic}. In the following an
overview is given of the most relevant \Rprog packages for forecasting at the
time of writing -- generally, the same functionalities are available in \Python packages.

Exponential smoothing models are popular and simple methods for time series. In
the exponential smoothing past observations are exponentially weighted down,
thus older observations have less impact than newer. The \textit{Holt-Winters}
procedure, where three smoothing constants are used to describe the variation in
time of the parameters, is one of the most famous exponential smoothing
methods. \Citet{doi:10.1287/mnsc.6.3.324} extended the double exponential
smoothing formulation by Holt to capture the seasonality. The
\code{HoltWinters()} function from the
\rpackage{stats}{https://stat.ethz.ch/R-manual/R-devel/library/stats/html/HoltWinters.html}
package estimates parameters of the \textit{Holt-Winters} procedure. The \rpackage{fable}{http://fable.tidyverts.org/} package \citep{fable} provides a
state-space framework to create exponential smoothing models in the function
\code{ETS()}. The function is based on the exponential smoothing framework
presented by \cite{hyndman2008forecasting}. The \rpackage{smooth}{https://cran.r-project.org/package=smooth}
package also provides methods for exponential smoothing.

The classical ARMAX models can be fitted with the \code{arima()} function from
the \rpackage{stats}{https://stat.ethz.ch/R-manual/R-devel/library/stats/html/00Index.html}
package and the \code{Arima()} function from the
\rpackage{forecast}{https://pkg.robjhyndman.com/forecast/} package
\citep{Hyndman2008} provides automatic model selection with \code{arima()}. \Rprog Packages like
\rpackage{marima}{https://CRAN.R-project.org/package=marima}
\citep{spliid1983fast},
\rpackage{KFAS}{https://cran.r-project.org/package=KFAS},
\rpackage{sysid}{https://cran.r-project.org/package=sysid} and
\rpackage{dlm}{https://cran.r-project.org/package=dlm}
\citep{dlm2010} can also be used for fitting ARMAX models. \Citet{spliid1983fast}
proposed a very fast and simple method for parameter estimation in large
multivariate ARMAX models with a pseudo-regression method that repeats the
regression estimation until it converges. The other packages represent time
series and regression models as state-space models and use a Kalman or Bayesian
filter to include exogenous variables in the model, and optimally reconstruct
and predict the states.

State-space modelling is frequently used to describe time series data from a
dynamical system, e.g.\ a falling body, see \citep{madsen2007time}. The
dynamical system can in such cases be written as differential equations or
difference equations. State-space models use filter techniques to optimally
reconstruct and predict the states, e.g.\ the Kalman filter, the extended Kalman
filter or other Bayesian filters. This gives the possibility of tracking the
coefficients over time, i.e.\ time-varying parameter estimation. The
\rpackage{KFAS}{https://cran.r-project.org/package=KFAS} package
\citep{kfas2017} provides state-space modelling, where the observations come
from the exponential family, e.g.\ Gaussian or Poisson. The
\rpackage{ctsm-r}{http://ctsm.info/} package provides a framework for
identifying and estimating partially observed continuous-discrete time state
space models, referred to as grey-box models. This modelling approach bridges
the gap between physical and statistical modelling using Stochastic Differential
Equations (SDEs) to model the system equations in continuous time and the
measurement equations in discrete time. Packages for discrete time state-space
modelling are:
\rpackage{dlm}{https://cran.r-project.org/package=dlm} for
Bayesian analysis of dynamic linear models,
\rpackage{MARSS}{https://nwfsc-timeseries.github.io/MARSS/} and
\rpackage{SSsimple}{https://cran.r-project.org/package=SSsimple}
for fitting multivariate state-space models.

For non-parametric time series models, the number of available packages is
growing rapidly.\ \rpackage{NTS}{https://cran.r-project.org/package=NTS}
provides simulation, estimation, prediction and identification for non-linear
time series data. It also includes threshold autoregressive models
(e.g.\ self-exciting threshold autoregressive models) and neural network
estimation. \rpackage{tsDyn}{https://cran.r-project.org/package=tsDyn}
provides methods for estimating non-parametric time series models, including
neural network estimation. Neural network, deep learning and machine learning
methods are available in \Rprog for most methods. Recurrent neural networks are in
the \rpackage{rnn}{http://qua.st/rnn}, the
\rpackage{keras}{https://keras.rstudio.com/} and
\rpackage{tensorflow}{https://github.com/rstudio/tensorflow} packages. Additive
time series models, where non-linear trends are fitted with seasonality patterns
are in \rpackage{prophet}{https://github.com/facebook/prophet}.

Some packages can be useful for forecast evaluation,
e.g.\ \rpackage{ForecastTB}{https://cran.r-project.org/package=ForecastTB}
presented in \citep{bokde2020forecasttb}. Packages like
\rpackage{forecastML}{https://cran.r-project.org/package=forecastML} and
\rpackage{modeltime}{https://CRAN.R-project.org/package=modeltime} provide
functionality that simplifies the process of multi-step-ahead forecasting with
standard machine learning algorithms. This purpose of handling multi-step-ahead
forecasts is also a key feature of the \onlineforecast package. The classical
time series models, such as ARMAX and Exponential Smoothing models, are mostly
optimal for modelling Linear Time Invariant (LTI) systems however most systems
are not LTI. Furthermore, since a model is always a simplification of reality,
optimal multi-step forecasting is often not possible with the classical models,
especially when using exogenous inputs. For optimal multi-step ahead forecasting
the models must be tuned for each horizon -- which is exactly what the
\onlineforecast package does.

\subsection{Implementation of onlineforecast}

The \onlineforecast package builds on an advanced model setup for
forecasting. This model setup was developed for applications such as forecasting
wind power \citep{nielsen2002prediction} and thermal loads in district heating
\citep{nielsen2006modelling}. The significance of the package is in the
``online'' term, indicating that the model is updated when new observation
becomes available -- recursively updating the coefficients and generating new
forecasts at every point in time.

The objective of the package is to make it easy to set up and optimize models
for generating online multi-step forecasts. The package contains functionalities
not directly available elsewhere:
\begin{itemize}
\item Use of forecasts, e.g.\ NWPs, as input to multi-step forecast models.
\item Application of non-linear models with non-parametric and
  coefficient varying techniques.
\item Optimal tuning of models for multi-step horizons.
\item Recursive estimation for tracking time-varying systems.
\end{itemize}
The package also provides a framework for handling data and setting up models,
which makes it easy to apply in a wide range of forecasting applications.

\new{A model is an approximation to the real world, thus it will always be a
  simplification and can never predict perfectly. One of the main challenges of
  identifying a good forecast model is to find the most informative input
  variables and the best structure of the model. The package provides
  functionality for defining, validating and selecting models in a systematic
  way.

To introduce the \onlineforecast models consider the simplest model with one
input. It's the linear model for the $k$'th horizon
\begin{align}
    Y_{t+k|t} = \beta_{0,k} + \beta_{1,k} u_{t+k|t} + \varepsilon_{t+k|t}
\end{align}
where $Y_{t+k|t}$ is the response variable and $u_{t+k|t}$ is the input
variable. The coefficients are $\beta_{0,k}$ and $\beta_{1,k}$, note that they
are subscripted with $k$ to indicate that they are estimated for each
horizon. The error $\varepsilon_{t+k|t}$ represents the difference between the
model prediction and the observed value for the $k$-step horizon. The
interpretation of the subscript notation $t+k|t$ on a variable, is that it's
the $k$-step prediction calculated using only past information at time $t$,
usually referred to either ``conditional on time $t$'' or ``given time $t$''.}

\new{The package offers to estimate the coefficients using either the Least Squares
(LS) or Recursive Least Squares (RLS) method. In the LS method, the coefficients
are constant, while the in RLS method the coefficients can change over time
\begin{align}
    Y_{t+k|t} = \beta_{0,k,t} + \beta_{1,k,t} u_{t+k|t} + \varepsilon_{t+k|t}
\end{align}
as indicated with the $t$ on the coefficients. This allows for tracking
changes occurring over time.}

The \new{package allows for easy definition of transformations and thus the
  possibility to fit non-linear models e.g.
\begin{align}
    Y_{t+k|t} = \beta_{0,k,t} + \beta_{1,k,t} f(u_{t+k|t};\alpha) + \varepsilon_{t+k|t}
\end{align}
where the function $f(u_{t+k|t};\alpha)$ is some non-linear function of the
input $u_{t+k|t}$ with parameter $\alpha$, e.g.\ a low pass filter on the outdoor
temperature to model building heat dynamics. The package sets up tuning of the
non-linear function parameters, e.g.\ if the parameter $\alpha$ determines the
degree of low-pass filtering it can be tuned with an optimizer to match the
system dynamics inherent in the data at hand.}

\begin{figure}
\centering
\begin{knitrout}
\definecolor{shadecolor}{rgb}{0.969, 0.969, 0.969}\color{fgcolor}
\includegraphics[width=1\linewidth]{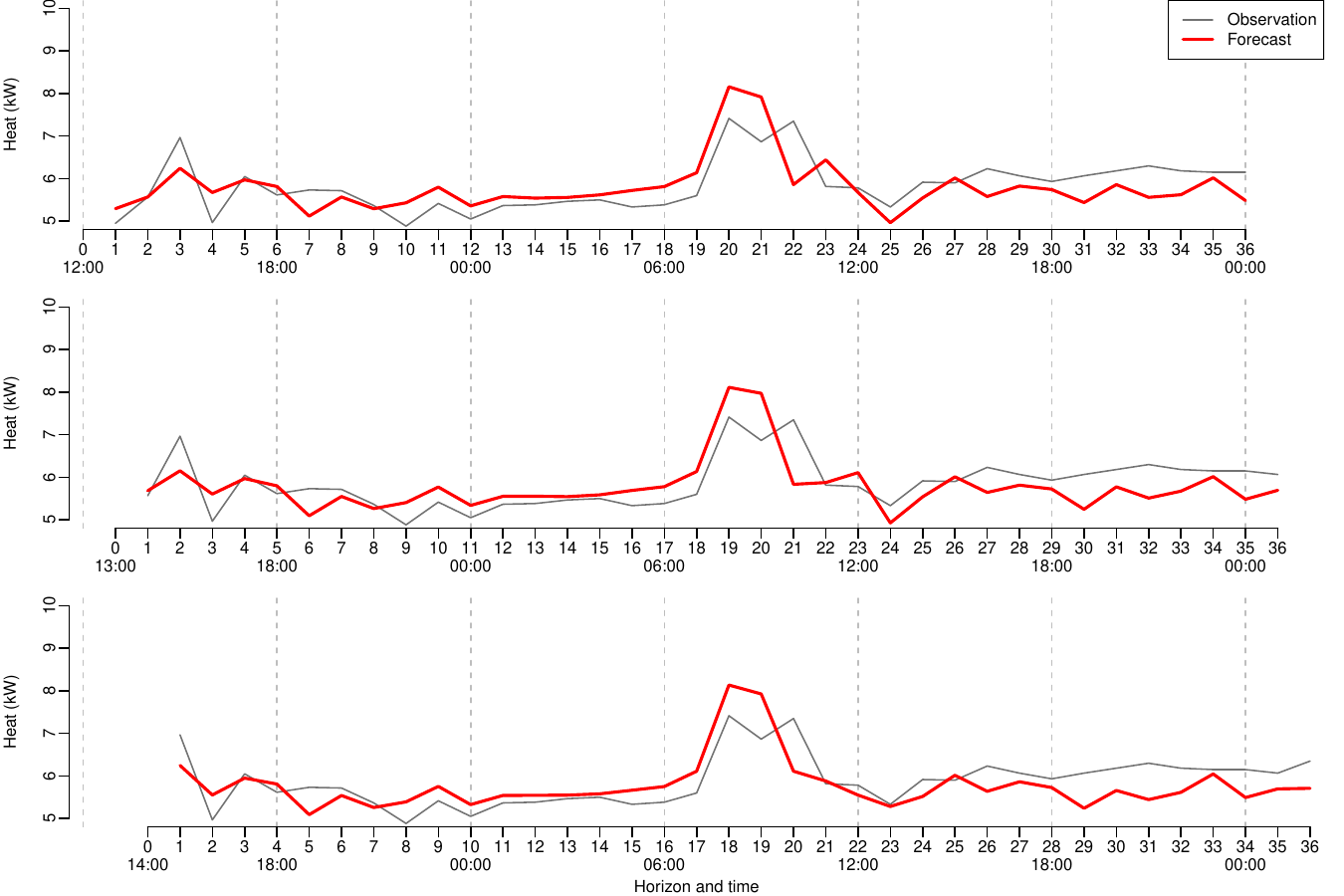} 
\end{knitrout}
\caption{Example of hourly load forecasts at three
  consecutive time steps. The upper is calculated at 12:00, the middle is
calculated at 13:00 and the lower at 14:00. It can be seen how the forecasts change
slightly as they are updated in each step, most clearly seen around 12:00 on day
2.}\label{fig:forecastsExample}
\end{figure}

An example of generated forecasts can be appreciated in Figure
\ref{fig:forecastsExample}. Hourly forecasts up to 36 steps ahead of heat load
in a single building are shown for three consecutive steps. This is the typical
structure of forecasts generated with the package. It can be seen how
the forecasts change slightly as they are updated in each step, e.g. around
12:00 at day 2, hence horizon $k=23$ in the upper plot, which corresponds to $k=21$ in the lower plot.

\FloatBarrier

\subsection{Vignettes}\label{subsec:vignettes}
A great way to get actual hands-on experience is through vignettes. They are
available when installing the package and on the website
\href{https://onlineforecasting.org}{onlineforecasting.org}, where also examples
of different forecast applications can be found. The package vignettes are:
\begin{itemize}
\item \vignette{setup-data} covers how data must be set up. The
  vignette goes into detail on how observations and model inputs
  (forecasts) are set up. The vignette also focuses on the importance of
  aligning forecasts correctly in time.
\item \vignette{setup-and-use-model} focus on how to set up a model and use it
  to generate forecasts.
\item \vignette{model-selection} demonstrates how model selection
  can be carried out.
\item \vignette{forecast-evaluation} covers the evaluation of forecasts, and how to
  use this information to improve a model.
\item \vignette{online-updating} demonstrates how to update a model in actual
  operation when new observations become available. This functionality isn't
  described in the \Rprog examples in the present paper.
\end{itemize}

Furthermore, one vignette is available only on the website:
\begin{itemize}
\item \vignette{nice-tricks} provides some useful tips on how to
  make the workflow easier with the package.
\end{itemize}

\subsection{Paper structure}

The structure of the paper is the following: In Section
\ref{sec:model-forecast-matrix} the notation used in the paper and how to set up
data is introduced. The core methodology is presented in Section
\ref{sec:two-stage-modelling} and important aspects of forecast modelling are
outlined in Section \ref{sec:model-select-valid}. In Section
\ref{sec:example-with-r-code} examples with \Rprog code are presented to provide a
short hands-on tutorial. The paper ends with a summary and conclusions in Section
\ref{sec:summary}.

In addition, three appendixes are included in the paper. In Appendix
\ref{sec:forec-model-notat} some guidelines on mathematical notation of
forecast models are provided. In Appendix \ref{app:transformation-of-inputs}
the functions used for transformations are detailed and in Appendix
\ref{app:regression} the regression schemes are covered in full detail.

\section{Notation and forecast matrices}\label{sec:model-forecast-matrix}
The notation in this article follows \cite{madsen2007time} as close as possible. All time series considered are equidistantly sampled and the sampling period is normalized to 1. Hence, the time $t$ is simply an integer indexing the value of a variable at time $t$. The same goes for $k$ which indexes the forecast horizon $k$ steps ahead. In the \onlineforecast setup, forecasts are calculated at time $t$ for each
horizon up to $\nk$ steps ahead. To achieve the desired notation that can
deal with overlapping time series, a two dimensional index is required. The
notation used is
\begin{align}
  u_{t+k|t}
\end{align}
which translates to: the value of variable $u$ at time $t+k$ \emph{conditional}
on the information available at time $t$. The conditional term is indicated
by the bar $|$. Thus, for $k > 0$ this is a forecast available at $t$ and $k$ is
the horizon. When writing a forecast model the following convention is used, here a simple example
\begin{align}
  Y_{t+k|t} = \beta_{0,k} + \beta_{1,k} u_{t+k|t} + \varepsilon_{t+k|t}
\end{align}
where $Y_{t+k|t}$ is the model output, $\beta_{0,k}$ and $\beta_{1,k}$ are the
coefficients and $\varepsilon_{t+k|t}$ with
$\mathrm{Var}(\varepsilon_{t+k|t})=\sigma^2_k$ is the error. The error process
and variance $\sigma^2_k$ is thus separate for each horizon. Note, that the
model is fitted separately for each horizon, so the coefficients take different
values for each horizon, and the predictions and errors are separated for each
horizon. This was a simplified example, see Appendix \ref{sec:forec-model-notat}
on how to write the full forecast models.

\subsection{Forecast matrix}

A forecast matrix is the format of forecast data in the onlineforecast setup. 
See examples in the \vignette{setup-data} vignette. Data
must have this format in order to be used as model input, and the forecasts
generated are in this format. The forecast matrix holds for any past time
\emph{the latest available forecast along the row} for the corresponding time
\begingroup
\setlength\arraycolsep{5pt} 
\renewcommand*{\arraystretch}{2} 
\begin{align}\label{eq:minput-def}
  \mB{u}_n = 
  \begin{blockarray}{cccccc}
    \textbf{k0} & \textbf{k1} & \textbf{k2} & \ldots & \textbf{k$n_k$} & \rightarrow\textbf{horizon/time}\downarrow\\
    \begin{block}{(ccccc)c}
      u_{1|1} & u_{2|1} & u_{3|1} & \ldots &
      u_{1+\nk|1} & 1\\    
      u_{2|2} & u_{3|2} & u_{4|2} & \ldots &
      u_{2+\nk|2} & 2\\    
      \vdots & \vdots & \vdots &  & \vdots & \vdots\\
      u_{t-1|t-1} & u_{t|t-1} & u_{t+1|t-1} &
      \ldots & u_{t-1+\nk|t-1} & t-1\\
      u_{t|t} & u_{t+1|t} & u_{t+2|t}  &
      \ldots & u_{t+\nk|t} & t\\
      \vdots & \vdots & \vdots &  & \vdots & \vdots\\
      u_{n|n} & u_{n+1|n} & u_{n+2|n}  &
      \ldots & u_{n+\nk|n} & n\\
    \end{block}
  \end{blockarray}
\end{align}
\endgroup
where 
\begin{itemize}
\item $t$ is the counter of time for equidistant time points with
  sampling period 1 (note that $t$ is not included in the matrix, it
  is simply the row number).
\item $n$ is the number of time points in the matrix. Hence, the data is
  available and can be used as model input at time $t=n$.
\item \nk is the longest forecasting horizon.
\item The column names (in \Rprog) are indicated above the matrix, they are simply
  a '{\bf k}' concatenated with the value of $k$, e.g.\ $n_k$ in the last column.
\end{itemize}

Note, that the $\textbf{k0}$ column holds values with forecast horizon $k=0$,
which could be real time observations. Usually, only the horizons to be
forecasted should be included, hence often $\textbf{k0}$ is not needed. For example with a prediction horizon $\nk=24$ at $t = 100$, we will have the
forecast matrix
\begingroup
\setlength\arraycolsep{5pt} 
\renewcommand*{\arraystretch}{2} 
\begin{align}
\mB{u}_{100} = 
\begin{blockarray}{cccccc}
  \textbf{k0} & \textbf{k1} & \textbf{k2} & \dots & \textbf{k24} & \rightarrow\textbf{horizon/time}\downarrow\\
  \begin{block}{(ccccc)c}
      u_{1|1} & u_{2|1} & u_{3|1} &
      \ldots & u_{25|1} & 1\\
      u_{2|2} & u_{3|2} & u_{4|2} &
      \ldots & u_{26|2} & 2\\
      \vdots & \vdots &  \vdots &  & \vdots & \vdots \\
      u_{99|99} & u_{100|99} & u_{101|99} &
      \ldots & u_{123|99} & 99\\
      u_{100|100} & u_{101|100} &
      u_{102|100}  & \ldots & u_{124|100} & 100\\
  \end{block}
\end{blockarray}
\end{align}
\endgroup
In Section \ref{sec:setup-of-data} examples of how data and forecast matrices are set up in \Rprog are given.

\section{Two-stage modelling procedure} \label{sec:two-stage-modelling}

A widespread approach to modelling non-linear functional relations between inputs
and output is a two-stage modelling procedure. See,
e.g.\ \cite{breiman1985estimating} and \cite{weisberg2005applied} for direct
transformation of predictor variables, and \cite{hastie2009elements} for
non-parametric transformation techniques (basis functions). Using
transformations allows for fitting complex models with robust and fast
estimation techniques. In the first stage, the \emph{transformation stage}, the inputs are mapped by
some function -- potentially into a higher dimensional space. In the second
stage, the \emph{regression stage}, a linear regression model\footnote{In the
remaining of the text, when the term ``regression'' is used it is implicit that
it's ``linear regression''.} is applied between the transformed inputs and the
output. An exemplification of this is presented in the following.

As an example a model with two inputs is presented. In this model the
transformation stage consists of generating an intercept and mapping the two
inputs (they are set up as forecast matrices $\mB{u}_{1,t}$ and $\mB{u}_{2,t}$)
\begin{align}\label{eq:transformation-stage-example}
  \text{Intercept:}&\quad x_{0,t+k|t} = 1 \\[1ex]
  \text{Input 1:}&\quad x_{1,t+k|t} = f_1(u_{1,t+k|t}, \mB{\alpha}\n{1})\\[1ex]
  \text{Input 2:}&\quad \mB{x}_{2,t+k|t} = f_2(u_{2,t+k|t}, \mB{\alpha}\n{2})
\end{align}
where the $f$'s are transformation functions that map the inputs to
regressors. Note, that the intercept is simply a constant passed on to the
regression. The transformations result in multiple inputs for the regression --
the latter actually as multiple variables indicated by the bold font
notation. In the regression stage the linear model
\begin{align}\label{eq:regression-stage-example}
  Y_{t+k|t} = \beta_{0,k} x_{0,t+k|t} + \beta_{1,k}
  x_{1,t+k|t} + \mB{\beta}^T_{2,k} \mB{x}_{2,t+k|t} + \varepsilon_{t+k|t}
\end{align}
is fitted. The regression is carried out separately for each horizon $k$. Thus, the combined model has:
\begin{itemize}
\item An intercept
\item Two inputs: $u_{1,t+k|t}$ and $u_{2,t+k|t}$
\item Output: $Y_{t+k|t}$
\item Transformation functions: $f_1$ and $f_2$
\item Transformation parameters: $\mB{\alpha}\n{1}$ and $\mB{\alpha}\n{2}$
\item Regression coefficients: $\beta_{0,k}$, $\beta_{1,k}$ and
  $\mB{\beta}_{2,k}$
\end{itemize}

Some transformation parameters should be optimized for the data at hand, e.g.\ a
low-pass filter coefficient depends on the system dynamics. The same goes for
some parameters related to the regression scheme, e.g.\ the forgetting factor
(introduced below). We will refer to them together as ``offline'' parameters.
The \onlineforecast package provides a setup where the offline parameters
can be optimized using a heuristic optimization (e.g., a BFGS quasi-Newton
method). The default score, which is minimized, is the Root Mean Square Error
(RMSE) of the predictions -- hence offline parameters in the
model above, given data from the period, $t = 1,2,\dots,n$, are found by
solving
\begin{align}\label{eq:rmse-score}
  \min_{\mB{\alpha}\n{1}, \mB{\alpha}\n{2}} \quad & \frac{1}{n-k}\sum_{t=1}^{n-k} (y_{t+k} -
  \hat{y}_{t+k|t}(\mB{\alpha}\n{1}, \mB{\alpha}\n{2}))^2
\end{align}
Naturally, other scores can be minimized (e.g.\ MAE or the Huber psi-function,
however the regression schemes should be modified accordingly, which is not
trivial).

The regression coefficients are calculated with a closed-form scheme: either
with the Least-Squares (LS) or the Recursive Least-Squares (RLS) scheme -- in
the latter the coefficients are allowed to vary over time. In both schemes the
coefficients are calculated separately for each horizon $k$. In Appendix
\ref{app:regression} both schemes are presented in full detail.

In the LS scheme the coefficients are gathered in the vector
$\hat{\mB{\beta}}_{k}$, which is constant in the entire period. The output
vector is $\mB{y}_{k,n}$ and for a given value of the transformation parameters
(i.e.\ here $\mB{\alpha}\n{1}$ and $\mB{\alpha}\n{2}$) the transformed data is
calculated and set up in the design matrix $\mB{X}_{k,n}$. The LS coefficients
are then calculated by
\begin{align}
  \hat{\mB{\beta}}_{k} = (\mB{X}_{k,n}\mB{X}_{k,n})^{-1} \mB{X}_{k,n}\mB{y}_{k,n}
\end{align}
and the predictions by
\begin{align}
  \hat{\mB{y}}_{k,n} = \mB{X}_{k,n} \hat{\mB{\beta}}_{k}
\end{align}
where $\hat{\mB{y}}_{k,n} = \left[ \hat{y}_{1+k|1} ~~
    \hat{y}_{2+k|2} ~~ \dots ~~ \hat{y}_{n|n-k}\right]^T$ are the
predictions. Note, that for the LS scheme the predictions are ``in-sample'',
since from the entire period was used for the coefficient estimation.

In the RLS scheme the coefficients are calculated recursively, meaning that they
are updated at each time $t$ when stepping through the period. In each update
only the ``newly'' obtained data at $t$ is used, thus only past data is used in
the calculations, which makes the calculated RLS coefficients and predictions
``out-of-sample'' (opposed to LS). Furthermore, the coefficients vary such that
the model adapts to the data over time. When writing the model as in Equation
\eqref{eq:regression-stage-example} this can be indicated with a $t$ subscript
on the coefficients, i.e.\ $\beta_{0,k,t}$, $\beta_{1,k,t}$ and
$\mB{\beta}_{2,k,t}$. The level of adaptivity can be controlled by setting the
forgetting factor $\lambda$ in the RLS scheme to a value between 0 and 1. For
$\lambda=1$ all past data at $t$ is equally weighted. For $\lambda < 1$ higher
weight is put on recent data -- the smaller the value the faster the model
adapts to data. By optimizing the forgetting factor as an offline parameter the
model adaptivity can be tuned to optimally track system changes over time.

An important point to notice is, that the offline parameters are always constant
for the given period, hence all predictions are essentially
``in-sample''. However, depending on the regression scheme there is a
difference: with the LS scheme the regression coefficients are constant through
the period, thus the predictions are (fully) ``in-sample'', where as with the
RLS scheme they adapt through the period and the predictions are
``out-of-sample'' (i.e.\ except for the offline parameters). This makes a
difference, since model over-fitting is less of a problem when using the RLS
scheme, although still problems can occur, e.g.\ if the forgetting factor is
optimized to be close to 1, because no systematic change occurred in the
training period, but changes appear in new data.

The typical \onlineforecast setup is to optimize the (usually few)
offline parameters in an ``offline'' setting, but calculate the
regression coefficients adaptively with the RLS. This has the advantage that the
model adapts and tracks the systematic changes in input-output relations, while
keeping the setup computationally very effective -- updating the coefficients
and calculating a forecast at each time $t$ takes few operations, since only the
newly available data is used. The more computationally heavy optimization of
offline parameters can be carried when computational resources are available
(e.g.\ every week for hourly forecasts).

In the following sections more information, on the package functions
for the two stages, are presented.

\subsection{Transformations stage}

In the transformation stage the inputs are mapped using some function as
demonstrated above, for more examples see the \vignette{setup-and-use-model}
vignette. The \onlineforecast package has functions available for most common
use, however it is easy to write and use new functions as they are simply \Rprog
functions, which return a forecast matrix (or a list of them), for more details
see Section \ref{sec:transformation-of-inputs}. The currently available
transformation functions are:

\begin{itemize}
\item Low-pass filtering, \code{lp()}: A low-pass filtering for modelling
  linear dynamics as a simple RC-model. See e.g.\ \cite{nielsen2006modelling}
  for further information.
\item Basis splines, \code{bspline()}: Use the \code{bs()} function for
  calculating regression splines basis functions.
\item Periodic basis splines, \code{pbspline()}: Use the \code{pbs()} function
  for calculating periodic regression splines basis functions.
\item Fourier series, \code{fs()}: Fourier series as periodic regression basis
  functions.
\item Auto-regressive, \code{AR()}: For including Auto-Regressive (AR) terms.
\item Intercept, \code{one()}: Generates a forecast matrix of ones, i.e.\ intercept. 
\end{itemize}

In the following subsection, the low-pass filtering is shortly described
below. In Appendix \ref{app:transformation-of-inputs} the other transformation
function are presented and in Section \ref{sec:example-with-r-code} a full
example with \Rprog code is provided.

The implementation in \onlineforecast allows all parameters, which are used in
some way (except the regression coefficients), to be included in an optimization
-- using any available optimizer i \Rprog. This includes e.g.\ the RLS forgetting
factor, knot points or order of splines -- hence both continuous and integer
variables. This functionality is achieved using a simple syntax as explained in
Section \ref{sec:example-with-r-code}.

\subsubsection{\it Low-pass filtering}
\label{sec:low-pass-filtering}

When modelling time series from linear dynamical systems, the classical ARMAX
model is often the optimal choice \citep{madsen2007time}. However, for
multi-step forecasting this is often not the case, especially for longer
horizons. In the \onlineforecast setup, where the regression model is fitted for
each horizon, a ``trick'' can be used for modelling linear dynamics: simply
apply a filter on the input and then use the filtered input in the regression
stage. For example, dynamics between ambient air temperature and indoor
temperature are slow due to the thermal mass of the building. Therefore, it is
reasonable that the dynamics between the ambient air temperature and the heat
demand (when keeping a desired indoor temperature) can be modelled using a
low-pass filter. This technique is successfully applied for energy modelling
using physical knowledge, see \cite{nielsen2006modelling} for modelling heat
load in district heating and \cite{bacher2013short} for forecasting single
buildings heat load.

In the package the simple low-pass filter
\begin{align}
  x_{t+k|t} = \frac{(1-a) u_{t+k|t}}{1-a x_{t-1+k|t-1}},
\end{align}
is implemented. The filter coefficient $a$ must take a value between 0 and 1 and
should be tuned to match the time constant optimal for the particular
data. When the current implemented low-pass filter is applied in the
transformation stage, on some forecast matrix $\mB{u}_{t+k|t}$, the filter is
applied on each column. Hence, independently for each horizon $k$. More advanced
filters can be implemented. See Appendix
\ref{sec:low-pass-filtering-appendix} for a more detailed description.

\subsection{Regression stage}

As described above and in full detail in Appendix \ref{app:regression}, the
regression model takes transformed data from a period $t = 1,2,\dots,n$ and is
fitted separately for each horizon $k$, i.e.\ the model structure remains the
same, only the coefficients change with the horizon. In the presented example, the regression model is as stated in Equation
\eqref{eq:regression-stage-example} -- it's implicit that the regression is
linear and that the coefficients are calculated using the LS or RLS scheme, both
are closed-form minimization of the $\mathit{RMSE}$ of the predictions. \new{The
  main differences between the two schemes have been outlined above and it
  should be clear that in most settings the RLS is preferable.}

Two fundamental assumptions are behind the optimality of least squares
predictions, hence the minimizing the $\mathit{RMSE}$. Both assumptions are
related to error process $\varepsilon_{t+k|t}$ in such models: \new{
\begin{itemize}
  \item The one-step error $\varepsilon_{t+1|t}$ is white noise, hence a
    sequence of independent, identically distributed (i.i.d.) random variables.
  \item They are drawn from a normal distribution, i.e.\ $\varepsilon_{t+k|t}
    \sim N(0,\sigma^2_k)$.
\end{itemize}}
The latter assumption is not very important, since first of all the LS ensures
the best and un-biased estimation of the conditional mean, which is often the
wanted and optimal point prediction \citep{madsen2007time}. For example, one
important feature is that sums of least squares predictions are also un-biased,
e.g.\ the daily sum of hourly LS predictions are good predictions of the daily
total. The i.i.d.\ assumption should be checked during model validation, as
described in the following section, with the Auto-Correlation Function (ACF) for
the one-step ahead residuals, as well as the Cross-Correlation Function (CCF)
between the one-step ahead residuals and the inputs, as demonstrated by
\cite{bacher2013short}.





\section{Model selection and validation} \label{sec:model-select-valid}

\subsection{Model selection}

In statistics, different model selection procedures are used
\citep{madsen2010introduction}. Essentially, a backward or a forward selection
procedure can be applied, or some combined approach. In the \onlineforecast
package both procedures are implemented, as well as a combined approach. The
following is a short description of the stepping process of each procedure
implemented, for examples of this see the \vignette{model-selection}
vignette. 

In each step of the selection process two properties of the model can be modified:
\begin{itemize}
\item Model inputs: In each step, inputs can either be removed or added.
\item Integer offline parameters: In each step integer parameters, such
  as the number of knot points in a basis spline or the number of harmonics in a
  Fourier series can be counted one up or down.
\end{itemize}

In each step of the process, the offline parameters are first optimized to
minimize the score for each modified model (in most cases the appropriate score
is the RMSE in Equation \eqref{eq:rmse-score} summed for selected horizons). Then the scores of the modified
models are compared with the score of the currently selected model and the model
with the lowest score is selected for the next step. This continues until no
further improvement of the score is achieved and the model with the lowest score is
selected. It's important to note, that the implemented procedure should only
be used with the RLS scheme, with the LS scheme the score is calculated fully
in-sample leading to over-fitting. For the LS an $F$-test should be applied
however that is currently not implemented.

A model must be given to initialize the stepping process. A backward selection
will start with the full model and one-by-one remove inputs and count down
parameters. A forward selection will start with the null model and, taking from
a provided full model, add inputs and count up parameters.  If the direction
``both'' is set then in each step, inputs will either be removed or added, and
parameters will be counted both up and down.

\subsection{Model validation}

The most important aspects of validation of forecast models are discussed in
this section, see the \vignette{forecast-evaluation} vignette for examples.

\subsubsection{\it Training and test set}
One fundamental caveat in data-driven modelling is over-fitting. This
can easily happen when the model is fitted (trained) and evaluated on the same
data. There are essentially two ways of dealing with this: Penalize increased
model complexity (regularization) or divide the data into a training set and
test set (cross-validation) \citep{tashman2000out}. In most forecasting applications the easiest and most transparent approach is
some cross-validation -- many methods for dividing into sets are possible. In
the \onlineforecast setup, when a model is fitted using a recursive estimation
method (like the RLS) only past data is used when calculating the regression
coefficients, so there is no need for dividing into a training set and a test
set.

The offline parameters (like the forgetting factor and low-pass filter
coefficients) are optimized on a particular period, hence over-fitting is
possible however it's most often very few parameters compared to the number of
observations -- so it's very unlikely to over-fit a recursive fitted model
in this setup. For non-recursive fitting, it is naturally important to divide into a training
and a test set -- which is easily done in the \onlineforecast setup using the
\code{scoreperiod} variable as demonstrated in Section
\ref{sec:example-with-r-code}.

\subsubsection{\it Scoring}
Scoring forecasts can be done in many ways, however in the \onlineforecast,
where the conditional mean is estimated and when using the RLS scheme,
it is straightforward to choose the Root Mean Square Error (RMSE) in Equation
\eqref{eq:rmse-score} as the best score to use. When using the LS scheme it can
be favourable to include regulation to avoid over-fitting, hence AIC or BIC
is preferable. One important point when comparing forecasts is to only include the complete
cases, i.e.\ forecasts at time points with no missing values across all horizons
and across all evaluated models. A function for easy selection of only complete
cases given multiple forecasts is implemented, see the examples in Section
\ref{subsec:evaluation}.

\subsubsection{\it Residual analysis}
Analysing the residuals is an important way to validate that a model cannot be
further improved or learn how it can improved. The main difference from
classical time series model validation, where only the one-step ahead error is
examined, is that multiple horizons should be included in the analysis. The two most important analysis:
\begin{itemize}
\item Plot residual time series to find where large forecast errors occur. The
  accumulated residuals are also often useful to examine, e.g.\ cumulative
  squared error plot.
\item Plot scatter plots of the residuals vs. other variables to see if any
  apparent dependencies are not described by the model.
\end{itemize}

In order to dig a bit more into the result of the recursive estimation, the
regression coefficients can be plotted over time. In this way, it is possible to
learn how the relations between the variables in the model evolve over time. If
drastic changes are found in some periods it might be worthwhile to zoom into
those periods to learn what causes these and potentially how to improve the
model. In case auto-correlation is left in the residuals, an error model can be
used to improve the forecasts by applying an auto-regressive model on the
residuals. This is somewhat equivalent to include an MA part in the original
model (which is not implemented as it is far from trivial).

As summarizing measures for validation of how well dynamics are modelled:
\begin{itemize}
\item Plot the auto-correlation function (ACF) of the one-step residuals.
\item Plot cross-correlation functions from one-step residuals to other
  variables, see \citep{bacher2013short}.
\end{itemize}
Systematic patterns found in these functions lead to direct knowledge on how to
improve the model, see for example the table on Page 155 in
\cite{madsen2007time} and the examples in Section \ref{subsec:evaluation} in the
present paper.

\section{Example with R code} \label{sec:example-with-r-code}

A short introduction to the basic functionalities and steps in setting up a
model is given in the following -- for more details and functionalities see the
vignettes listed in Section \ref{subsec:vignettes} and the website \href{https://onlineforecasting.org}{onlineforecasting.org}.

\subsection{Setup of data} \label{sec:setup-of-data}

As input to the model, we will use weather forecasts, which are arranged in forecast matrices, e.g.\ the ambient air temperature
\begingroup
\setlength\arraycolsep{5pt} 
\renewcommand*{\arraystretch}{2} 
\begin{align}
  \mBTa{n} = 
  \begin{blockarray}{cccccc}
    \textbf{k0} & \textbf{k1} & \textbf{k2} & \ldots & \textbf{kxx} & \rightarrow\textbf{horizon/time}\downarrow\\
    \begin{block}{(ccccc)c}
      \Ta{1|1} & \Ta{2|1} & \Ta{3|1} & \ldots &
      \Ta{1+\nk|1} & 1\\    
      \Ta{2|2} & \Ta{3|2} & \Ta{4|2} & \ldots &
      \Ta{2+\nk|2} & 2\\    
      \vdots & \vdots & \vdots &  & \vdots & \vdots\\
      \Ta{n-1|n-1} & \Ta{n|n-1} & \Ta{n+1|n-1} &
      \ldots & \Ta{n-1+\nk|n-1} & n-1\\
      \Ta{n|n} & \Ta{n+1|n} & \Ta{n+2|n}  &
      \ldots & \Ta{n+\nk|n} & n\\
    \end{block}
  \end{blockarray}
\end{align}
\endgroup
Building heat load and weather forecast data is available in \code{Dbuilding}
after loading the package:
\begin{knitrout}
\definecolor{shadecolor}{rgb}{0.969, 0.969, 0.969}\color{fgcolor}\begin{kframe}
\begin{alltt}
\hlstd{D} \hlkwb{<-} \hlstd{Dbuilding}
\hlkwd{str}\hlstd{(D,} \hlnum{1}\hlstd{)}
\end{alltt}
\begin{verbatim}
## List of 7
##  $ t            : POSIXct[1:1824], format: "2010-12-15 01:00:00" ...
##  $ heatload     : num [1:1824] 5.92 5.85 5.85 5.88 5.85 ...
##  $ heatloadtotal: num [1:1824] 4 4.07 4.08 4.01 4.46 ...
##  $ Taobs        : num [1:1824] -6.19 -5.7 -5.05 -5.1 -5.62 ...
##  $ Iobs         : num [1:1824] 0 0 5.41 12.99 13.01 ...
##  $ Ta           :'data.frame':	1824 obs. of  36 variables:
##  $ I            :'data.frame':	1824 obs. of  36 variables:
##  - attr(*, "class")= chr [1:2] "data.list" "list"
\end{verbatim}
\begin{alltt}
\hlkwd{class}\hlstd{(D)}
\end{alltt}
\begin{verbatim}
## [1] "data.list" "list"
\end{verbatim}
\end{kframe}
\end{knitrout}
\noindent The \code{data.list} class comes with the package, it's actually a
\code{list} with some predefined structure, see \code{?data.list}.

\noindent All inputs to be used in a model must be given as forecast matrices. The ambient
temperature forecast is set up in a forecast matrix, which class is just a
\code{data.frame}:
\begin{knitrout}
\definecolor{shadecolor}{rgb}{0.969, 0.969, 0.969}\color{fgcolor}\begin{kframe}
\begin{alltt}
\hlkwd{class}\hlstd{(D}\hlopt{$}\hlstd{Ta)}
\end{alltt}
\begin{verbatim}
## [1] "data.frame"
\end{verbatim}
\begin{alltt}
\hlkwd{head}\hlstd{(D}\hlopt{$}\hlstd{Ta[ ,}\hlnum{1}\hlopt{:}\hlnum{8}\hlstd{],} \hlnum{4}\hlstd{)}
\end{alltt}
\begin{verbatim}
##         k1       k2      k3      k4       k5       k6       k7       k8
## 1 -2.82340 -3.20275 -3.1185 -3.0896 -3.13200 -3.16130 -3.16645 -3.08885
## 2 -2.90405 -3.11850 -3.0896 -3.1320 -3.16130 -3.16645 -3.08885 -2.77165
## 3 -2.93590 -3.08960 -3.1320 -3.1613 -3.16645 -3.08885 -2.77165 -2.32185
## 4 -2.89315 -3.11285 -3.0484 -3.1090 -3.11600 -2.80990 -2.36895 -2.00945
\end{verbatim}
\end{kframe}
\end{knitrout}
\noindent The time is kept in the vector $t$ where the time stamps are set in the end of the hour:
\begin{knitrout}
\definecolor{shadecolor}{rgb}{0.969, 0.969, 0.969}\color{fgcolor}\begin{kframe}
\begin{alltt}
\hlstd{D}\hlopt{$}\hlstd{t[}\hlnum{1}\hlopt{:}\hlnum{4}\hlstd{]}
\end{alltt}
\begin{verbatim}
## [1] "2010-12-15 01:00:00 UTC" "2010-12-15 02:00:00 UTC"
## [3] "2010-12-15 03:00:00 UTC" "2010-12-15 04:00:00 UTC"
\end{verbatim}
\end{kframe}
\end{knitrout}
\noindent Observations are simply in a vector:
\begin{knitrout}
\definecolor{shadecolor}{rgb}{0.969, 0.969, 0.969}\color{fgcolor}\begin{kframe}
\begin{alltt}
\hlstd{D}\hlopt{$}\hlstd{heatload[}\hlnum{1}\hlopt{:}\hlnum{4}\hlstd{]}
\end{alltt}
\begin{verbatim}
## [1] 5.916667 5.850000 5.850000 5.883333
\end{verbatim}
\end{kframe}
\end{knitrout}
\noindent For more details on the \code{data.list} class, see the
\href{https://onlineforecasting.org/vignettes/setup-data.html}{setup-data}
vignette with \code{vignette("setup-data")} -- which demonstrates useful
functions for manipulating and exploring forecast data.

\subsection{Defining a model} \label{sec:defining-model}

Models are set up using the R6 class \code{forecastmodel}. An object
of the class is instantiated by:
\begin{knitrout}
\definecolor{shadecolor}{rgb}{0.969, 0.969, 0.969}\color{fgcolor}\begin{kframe}
\begin{alltt}
\hlstd{model} \hlkwb{<-} \hlstd{forecastmodel}\hlopt{$}\hlkwd{new}\hlstd{()}
\end{alltt}
\end{kframe}
\end{knitrout}
\noindent It holds variables and functions for representing and manipulating a model.

\noindent We want to forecast the \code{heatload} variable in the data list, so we set that as
the model output by:
\begin{knitrout}
\definecolor{shadecolor}{rgb}{0.969, 0.969, 0.969}\color{fgcolor}\begin{kframe}
\begin{alltt}
\hlstd{model}\hlopt{$}\hlstd{output} \hlkwb{<-} \hlstr{"heatload"}
\end{alltt}
\end{kframe}
\end{knitrout}
\noindent The model inputs and transformations must then be defined. We can add
an input as a linear function by:
\begin{knitrout}
\definecolor{shadecolor}{rgb}{0.969, 0.969, 0.969}\color{fgcolor}\begin{kframe}
\begin{alltt}
\hlstd{model}\hlopt{$}\hlkwd{add_inputs}\hlstd{(}\hlkwc{Ta} \hlstd{=} \hlstr{"Ta"}\hlstd{)}
\end{alltt}
\end{kframe}
\end{knitrout}
\noindent hence the name of the ambient temperature forecast matrix. Then the $k$ horizon
column becomes $\beta_k T\ns{a}{t+k|t}$ in the regression for the $k$ horizon.

\noindent Adding an intercept to a model can be done by:
\begin{knitrout}
\definecolor{shadecolor}{rgb}{0.969, 0.969, 0.969}\color{fgcolor}\begin{kframe}
\begin{alltt}
\hlstd{model}\hlopt{$}\hlkwd{add_inputs}\hlstd{(}\hlkwc{mu} \hlstd{=} \hlstr{"one()"}\hlstd{)}
\end{alltt}
\end{kframe}
\end{knitrout}
\noindent where the function \code{one()} simply returns a vector of 1's, which will be
inserted in the design matrix, see in Appendix \ref{app:regression}. The
function is run during the transformation, which is carried out with the function \code{model\$transform\_data()}. Most of the time, \code{model\$transform\_data()} is called inside a ``fitting function'' when a model is fitted.

\subsubsection{\it Input transformation} \label{sec:transformation-of-inputs}

Transformations (or mappings) of inputs are simply \Rprog code which returns
a forecast matrix or a list of forecast matrices. Dynamics can be modelled using filters. For example, low-pass filtering of a variable with:
\begin{knitrout}
\definecolor{shadecolor}{rgb}{0.969, 0.969, 0.969}\color{fgcolor}\begin{kframe}
\begin{alltt}
\hlstd{model}\hlopt{$}\hlkwd{add_inputs}\hlstd{(}\hlkwc{Ta} \hlstd{=} \hlstr{"lp(Ta, a1=0.9)"}\hlstd{)}
\end{alltt}
\end{kframe}
\end{knitrout}
\noindent will, when the transformation is run, apply a low-pass filter along each column
of the forecast matrix \code{Ta}. The filter coefficient is set to $a=0.9$.
To illustrate the effect of this, see Appendix
\ref{app:transformation-of-inputs} and the vignette
\href{https://onlineforecasting.org/vignettes/setup-and-use-model.html#input-transformations}{setup-and-use-model}. Non-linear effects can be modelled using basis functions. For mapping an input to basis splines the function \code{bspline()} is
provided. It's a wrapper of the \code{bs()} function from the \rpackage{splines}{https://www.rdocumentation.org/search?q=splines} package and has the same arguments. To e.g.\ include a non-linear function of the ambient temperature:
\begin{knitrout}
\definecolor{shadecolor}{rgb}{0.969, 0.969, 0.969}\color{fgcolor}\begin{kframe}
\begin{alltt}
\hlstd{model}\hlopt{$}\hlkwd{add_inputs}\hlstd{(}\hlkwc{Ta} \hlstd{=} \hlstr{"bspline(Ta, df=5)"}\hlstd{)}
\end{alltt}
\end{kframe}
\end{knitrout}
\noindent where \code{df} is the degrees of freedom of the spline function.

\noindent Functions can be nested, e.g.\ first a low-pass filter before mapping to basis
splines:
\begin{knitrout}
\definecolor{shadecolor}{rgb}{0.969, 0.969, 0.969}\color{fgcolor}\begin{kframe}
\begin{alltt}
\hlstd{model}\hlopt{$}\hlkwd{add_inputs}\hlstd{(}\hlkwc{Ta} \hlstd{=} \hlstr{"bspline(lp(Ta, a1=0.9), df=5)"}\hlstd{)}
\end{alltt}
\end{kframe}
\end{knitrout}

\noindent Fourier series can also be used as basis functions. They can be generated with the \code{fs()} function, e.g.\ if \code{tday} is a forecast matrix with the time of day in hours (range from 0 to 24) then:
\begin{knitrout}
\definecolor{shadecolor}{rgb}{0.969, 0.969, 0.969}\color{fgcolor}\begin{kframe}
\begin{alltt}
\hlstd{model}\hlopt{$}\hlkwd{add_inputs}\hlstd{(}\hlkwc{mutday} \hlstd{=} \hlstr{"fs(tday/24, nharmonics=10)"}\hlstd{)}
\end{alltt}
\end{kframe}
\end{knitrout}
\noindent generates the Fourier series for fitting a diurnal curve with 10 harmonics
(since there is both a cosine and a sine for each harmonic it results in
20 inputs in the regression model).

\new{Varying-coefficient models can be realized with multiplication of inputs
\citep{hastie1993varying}.} The multiplication operator \code{\%**\%} must be used, see how with
\code{?\%**\%}. For example, models similar to those proposed by
\citet{rasmussen2020semi}, where the coefficient for global radiation change as
a function of the time of day \code{tday}, can be realised by:
\begin{knitrout}
\definecolor{shadecolor}{rgb}{0.969, 0.969, 0.969}\color{fgcolor}\begin{kframe}
\begin{alltt}
\hlstd{model}\hlopt{$}\hlkwd{add_inputs}\hlstd{(}\hlkwc{I} \hlstd{=} \hlstr{"bspline(tday, df=5) %**% I"}\hlstd{)}
\end{alltt}
\end{kframe}
\end{knitrout}
\noindent For more details, see the
\href{https://onlineforecasting.org/examples/solar-power-forecasting.html}{solar-power-forecasting}
example on the website.

\noindent Finally, it's often useful to add an auto-regressive input. This can be done by:
\begin{knitrout}
\definecolor{shadecolor}{rgb}{0.969, 0.969, 0.969}\color{fgcolor}\begin{kframe}
\begin{alltt}
\hlstd{model}\hlopt{$}\hlkwd{add_inputs}\hlstd{(}\hlkwc{AR} \hlstd{=} \hlstr{"AR(c(0,1))"}\hlstd{)}
\end{alltt}
\end{kframe}
\end{knitrout}
\noindent which adds the output observations at time $t$ and time $t-1$ as regression inputs.

\subsection{Model fitting and parameter optimization}

After setting up a model it can be fitted to data. This implies carrying out the
transformation and regression stages, which is done by passing the model to a
fitting function. Different functions implement different regression schemes. Both the two currently available fitting functions
\code{lm\_fit()} and \code{rls\_fit()} takes a the offline parameters as a
vector, fits a model and returns the RMSE (summed for all horizons).

\noindent To demonstrate this we replace the inputs on the previous defined model with two inputs:
\begin{knitrout}
\definecolor{shadecolor}{rgb}{0.969, 0.969, 0.969}\color{fgcolor}\begin{kframe}
\begin{alltt}
\hlstd{model}\hlopt{$}\hlstd{inputs} \hlkwb{<-} \hlkwa{NULL}
\hlstd{model}\hlopt{$}\hlkwd{add_inputs}\hlstd{(}\hlkwc{mu} \hlstd{=} \hlstr{"one()"}\hlstd{,}
                 \hlkwc{Ta} \hlstd{=} \hlstr{"lp(Ta, a1=0.9)"}\hlstd{)}
\end{alltt}
\end{kframe}
\end{knitrout}
\noindent We also have to set the ``score period'', which is simply a logical vector that
specifies the time points to be included in the score calculation. For the linear regression we can include all points by:
\begin{knitrout}
\definecolor{shadecolor}{rgb}{0.969, 0.969, 0.969}\color{fgcolor}\begin{kframe}
\begin{alltt}
\hlstd{D}\hlopt{$}\hlstd{scoreperiod} \hlkwb{<-} \hlkwd{rep}\hlstd{(}\hlnum{TRUE}\hlstd{,} \hlkwd{length}\hlstd{(D}\hlopt{$}\hlstd{t))}
\end{alltt}
\end{kframe}
\end{knitrout}
\noindent We can now fit the model and obtain the summed RMSE for the horizons 1 to 6 steps ahead by:
\begin{knitrout}
\definecolor{shadecolor}{rgb}{0.969, 0.969, 0.969}\color{fgcolor}\begin{kframe}
\begin{alltt}
\hlstd{model}\hlopt{$}\hlstd{kseq} \hlkwb{<-} \hlnum{1}\hlopt{:}\hlnum{6}
\hlkwd{lm_fit}\hlstd{(}\hlkwd{c}\hlstd{(}\hlkwc{Ta__a1}\hlstd{=}\hlnum{0.8}\hlstd{), model, D,} \hlkwc{scorefun}\hlstd{=rmse,} \hlkwc{returnanalysis}\hlstd{=}\hlnum{FALSE}\hlstd{)}
\end{alltt}
\begin{verbatim}
## [1] 5.039611
\end{verbatim}
\end{kframe}
\end{knitrout}
\noindent The function can be passed to an optimizer, which can then find the
parameter value(s) which minimize the score.

\noindent To facilitate the optimization wrappers for \code{optim()} are included. We define parameter(s) to optimize:
\begin{knitrout}
\definecolor{shadecolor}{rgb}{0.969, 0.969, 0.969}\color{fgcolor}\begin{kframe}
\begin{alltt}
\hlstd{model}\hlopt{$}\hlkwd{add_prmbounds}\hlstd{(}\hlkwc{Ta__a1} \hlstd{=} \hlkwd{c}\hlstd{(}\hlkwc{min}\hlstd{=}\hlnum{0.8}\hlstd{,} \hlkwc{init}\hlstd{=}\hlnum{0.9}\hlstd{,} \hlkwc{max}\hlstd{=}\hlnum{0.9999}\hlstd{))}
\end{alltt}
\end{kframe}
\end{knitrout}
\noindent Note, the double underscore syntax. The double underscore separates the input
name and the name of the parameter. So in the case above the value of \code{a1}
in the \Rprog code for input \code{Ta} will be optimized, starting at an initial
value of 0.9 and staying within the specified limits.

We can then run the optimization calculating scores (to save computational time
run on only a horizons 3 and 18 steps ahead):
\begin{knitrout}
\definecolor{shadecolor}{rgb}{0.969, 0.969, 0.969}\color{fgcolor}\begin{kframe}
\begin{alltt}
\hlkwd{lm_optim}\hlstd{(model, D,} \hlkwc{kseq}\hlstd{=}\hlkwd{c}\hlstd{(}\hlnum{3}\hlstd{,}\hlnum{18}\hlstd{))}
\end{alltt}
\end{kframe}
\end{knitrout}
\noindent The \code{lm\_optim()} function is a wrapper for the \Rprog optimizer function
\code{optim()}. It returns the result from \code{optim()} and sets optimized
parameters in:
\begin{knitrout}
\definecolor{shadecolor}{rgb}{0.969, 0.969, 0.969}\color{fgcolor}\begin{kframe}
\begin{alltt}
\hlstd{model}\hlopt{$}\hlstd{prm}
\end{alltt}
\begin{verbatim}
##    Ta__a1 
## 0.8926342
\end{verbatim}
\end{kframe}
\end{knitrout}

\noindent If we want to carry out the regression using the RLS scheme,
we must have a burn-in period, i.e.\ exclude a period in the beginning of the
data from the calculation of the score, e.g.\ 5 days:
\begin{knitrout}
\definecolor{shadecolor}{rgb}{0.969, 0.969, 0.969}\color{fgcolor}\begin{kframe}
\begin{alltt}
\hlstd{D}\hlopt{$}\hlstd{scoreperiod} \hlkwb{<-} \hlkwd{in_range}\hlstd{(}\hlkwd{min}\hlstd{(D}\hlopt{$}\hlstd{t)}\hlopt{+}\hlnum{5}\hlopt{*}\hlnum{24}\hlopt{*}\hlnum{3600}\hlstd{, D}\hlopt{$}\hlstd{t)}
\end{alltt}
\end{kframe}
\end{knitrout}
\noindent and we have to set the forgetting factor as a regression parameter:
\begin{knitrout}
\definecolor{shadecolor}{rgb}{0.969, 0.969, 0.969}\color{fgcolor}\begin{kframe}
\begin{alltt}
\hlstd{model}\hlopt{$}\hlkwd{add_regprm}\hlstd{(}\hlstr{"rls_prm(lambda=0.9)"}\hlstd{)}
\end{alltt}
\end{kframe}
\end{knitrout}

\noindent We can now optimize with RLS:
\begin{knitrout}
\definecolor{shadecolor}{rgb}{0.969, 0.969, 0.969}\color{fgcolor}\begin{kframe}
\begin{alltt}
\hlkwd{rls_optim}\hlstd{(model, D,} \hlkwc{kseq}\hlstd{=}\hlkwd{c}\hlstd{(}\hlnum{3}\hlstd{,}\hlnum{18}\hlstd{))}
\end{alltt}
\end{kframe}
\end{knitrout}
\noindent and if we want to also optimize the forgetting factor:
\begin{knitrout}
\definecolor{shadecolor}{rgb}{0.969, 0.969, 0.969}\color{fgcolor}\begin{kframe}
\begin{alltt}
\hlstd{model}\hlopt{$}\hlkwd{add_prmbounds}\hlstd{(}\hlkwc{lambda} \hlstd{=} \hlkwd{c}\hlstd{(}\hlkwc{min}\hlstd{=}\hlnum{0.7}\hlstd{,} \hlkwc{init}\hlstd{=}\hlnum{0.99}\hlstd{,} \hlkwc{max}\hlstd{=}\hlnum{0.9999}\hlstd{))}
\hlkwd{rls_optim}\hlstd{(model, D,} \hlkwc{kseq}\hlstd{=}\hlkwd{c}\hlstd{(}\hlnum{3}\hlstd{,}\hlnum{18}\hlstd{))}
\end{alltt}
\end{kframe}
\end{knitrout}

\subsubsection{\it Calculating forecasts}

While developing models it's most convenient to use the fit functions for
calculating predictions, e.g.: 
\begin{knitrout}
\definecolor{shadecolor}{rgb}{0.969, 0.969, 0.969}\color{fgcolor}\begin{kframe}
\begin{alltt}
\hlstd{model}\hlopt{$}\hlstd{kseq} \hlkwb{<-} \hlnum{1}\hlopt{:}\hlnum{24}
\hlstd{fit} \hlkwb{<-} \hlkwd{rls_fit}\hlstd{(model}\hlopt{$}\hlstd{prm, model, D)}
\end{alltt}
\end{kframe}
\end{knitrout}
\noindent will return a list holding the forecasts (in the forecast matrix \code{fit\$Yhat}) and other useful information. Forecasts can also be calculated directly with a predict function:
\begin{knitrout}
\definecolor{shadecolor}{rgb}{0.969, 0.969, 0.969}\color{fgcolor}\begin{kframe}
\begin{alltt}
\hlkwd{rls_predict}\hlstd{(model, model}\hlopt{$}\hlkwd{transform_data}\hlstd{(D))}
\end{alltt}
\end{kframe}
\end{knitrout}
\noindent will return a forecast matrix using the input data in \code{D}.

\subsection{Evaluation}\label{subsec:evaluation}

Finally, it's time to evaluate the forecasts and potentially get inspired to
improve the model. For a comprehensive introduction, see the
\vignette{forecast-evaluation} vignette.

\noindent First, a plot of the forecasts is always a good idea to learn how the model
work:
\begin{knitrout}
\definecolor{shadecolor}{rgb}{0.969, 0.969, 0.969}\color{fgcolor}\begin{kframe}
\begin{alltt}
\hlstd{D}\hlopt{$}\hlstd{Yhat} \hlkwb{<-} \hlstd{fit}\hlopt{$}\hlstd{Yhat}
\hlkwd{plot_ts}\hlstd{(}\hlkwd{subset}\hlstd{(D,D}\hlopt{$}\hlstd{scoreperiod),} \hlstr{"heatload$|Yhat"}\hlstd{,} \hlkwc{kseq}\hlstd{=}\hlkwd{c}\hlstd{(}\hlnum{1}\hlstd{,}\hlnum{5}\hlstd{,}\hlnum{24}\hlstd{),} \hlkwc{p}\hlstd{=p)}
\end{alltt}
\end{kframe}
\includegraphics[width=1\linewidth]{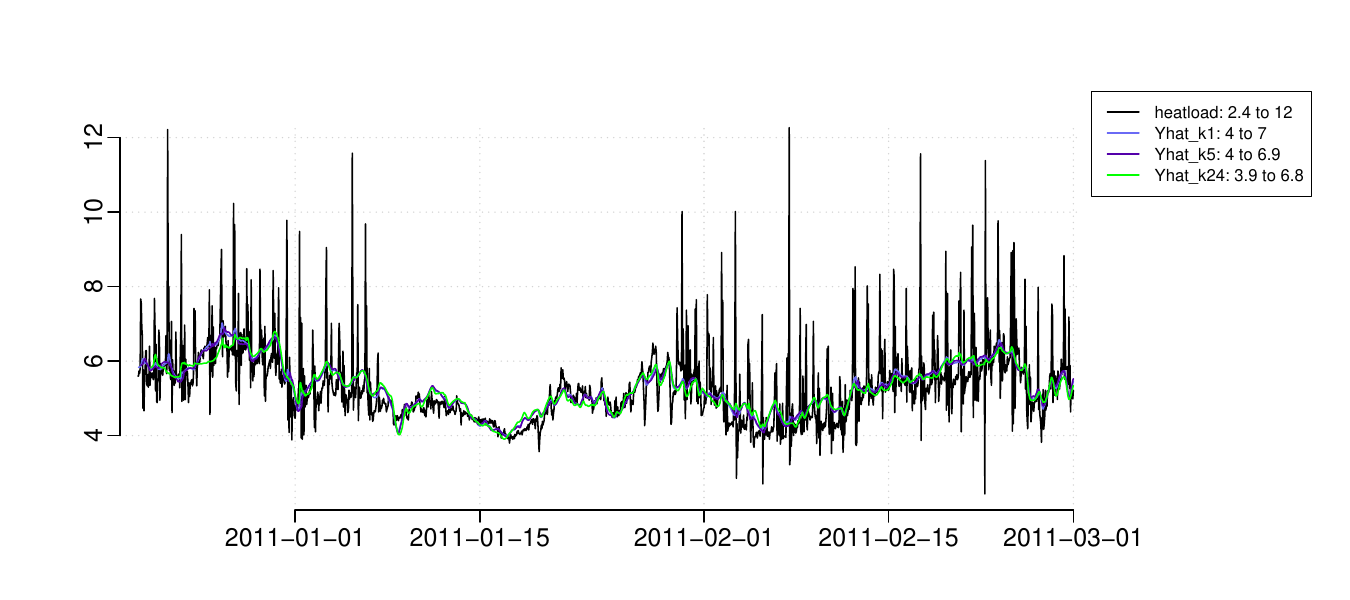} 
\end{knitrout}

\noindent We can extract the residuals from the fit:
\begin{knitrout}
\definecolor{shadecolor}{rgb}{0.969, 0.969, 0.969}\color{fgcolor}\begin{kframe}
\begin{alltt}
\hlstd{R} \hlkwb{<-} \hlkwd{residuals}\hlstd{(fit)}
\hlstd{R[}\hlnum{101}\hlopt{:}\hlnum{105}\hlstd{,} \hlnum{1}\hlopt{:}\hlnum{5}\hlstd{]}
\end{alltt}
\begin{verbatim}
##              h1           h2           h3           h4           h5
## 101 -0.30889122 -0.311105984 -0.311287844 -0.317092662 -0.309276247
## 102 -0.15452158 -0.168868282 -0.168640790 -0.171378395 -0.176755885
## 103  0.02755442  0.008792884  0.007182829  0.004900251  0.001789956
## 104  1.78881155  1.768225483  1.768757817  1.763675174  1.762632322
## 105  2.38451273  2.394916757  2.398424406  2.394625320  2.389784248
\end{verbatim}
\end{kframe}
\end{knitrout}
\noindent Note, that the column names are with \code{h} prefix. Hence, it's not a forecast
matrix, since each column is a time series aligned with time $t$. They are observations of the error.

\noindent Plot the ACF of the one-step residuals:
\begin{knitrout}
\definecolor{shadecolor}{rgb}{0.969, 0.969, 0.969}\color{fgcolor}\begin{kframe}
\begin{alltt}
\hlkwd{acf}\hlstd{(R}\hlopt{$}\hlstd{h1,} \hlkwc{na.action}\hlstd{=na.pass,} \hlkwc{lag.max}\hlstd{=}\hlnum{96}\hlstd{,} \hlkwc{main}\hlstd{=}\hlstr{""}\hlstd{)}
\end{alltt}
\end{kframe}
\includegraphics[width=1\linewidth]{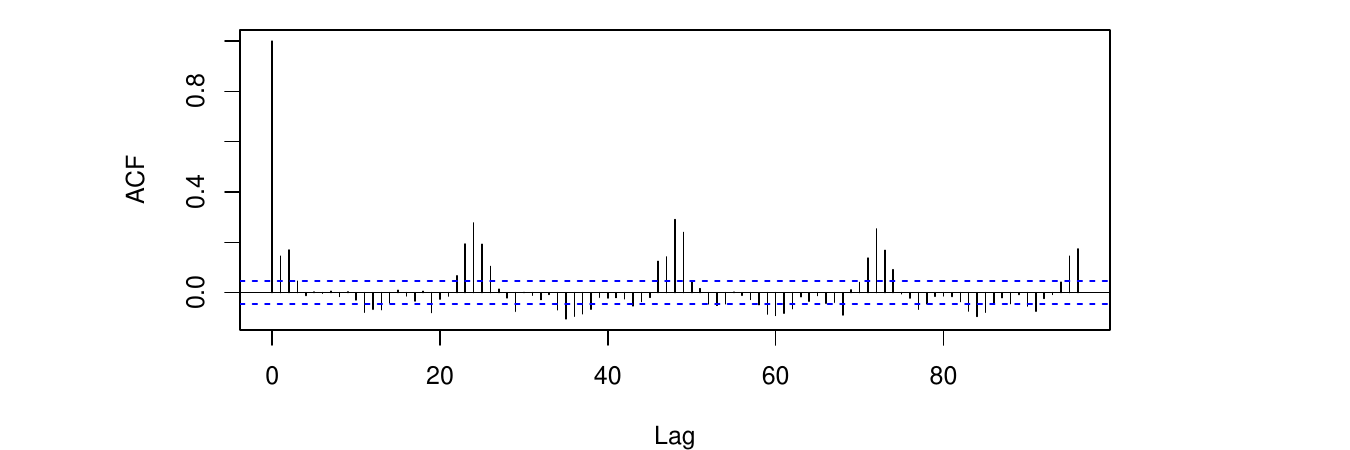} 
\end{knitrout}

\noindent The ACF plot suggest that there remains a diurnal pattern to be modelled. It can
be achieve by adding a diurnal curve to the model, e.g.\ with Fourier series
basis functions. This is demonstrated in the vignette \vignette{setup-and-use-model}.

\noindent We also want to calculate the score as a function of the horizon:
\begin{knitrout}
\definecolor{shadecolor}{rgb}{0.969, 0.969, 0.969}\color{fgcolor}\begin{kframe}
\begin{alltt}
\hlstd{inscore} \hlkwb{<-} \hlstd{D}\hlopt{$}\hlstd{scoreperiod} \hlopt{&} \hlkwd{complete_cases}\hlstd{(fit}\hlopt{$}\hlstd{Yhat)}
\hlstd{RMSE} \hlkwb{<-} \hlkwd{score}\hlstd{(}\hlkwd{residuals}\hlstd{(fit),} \hlkwc{scoreperiod} \hlstd{= inscore)}
\hlkwd{plot}\hlstd{(RMSE,} \hlkwc{ylim}\hlstd{=}\hlkwd{c}\hlstd{(}\hlnum{0.75}\hlstd{,}\hlnum{0.88}\hlstd{),} \hlkwc{xlab}\hlstd{=}\hlstr{"Horizon"}\hlstd{)}
\end{alltt}
\end{kframe}
\includegraphics[width=1\linewidth]{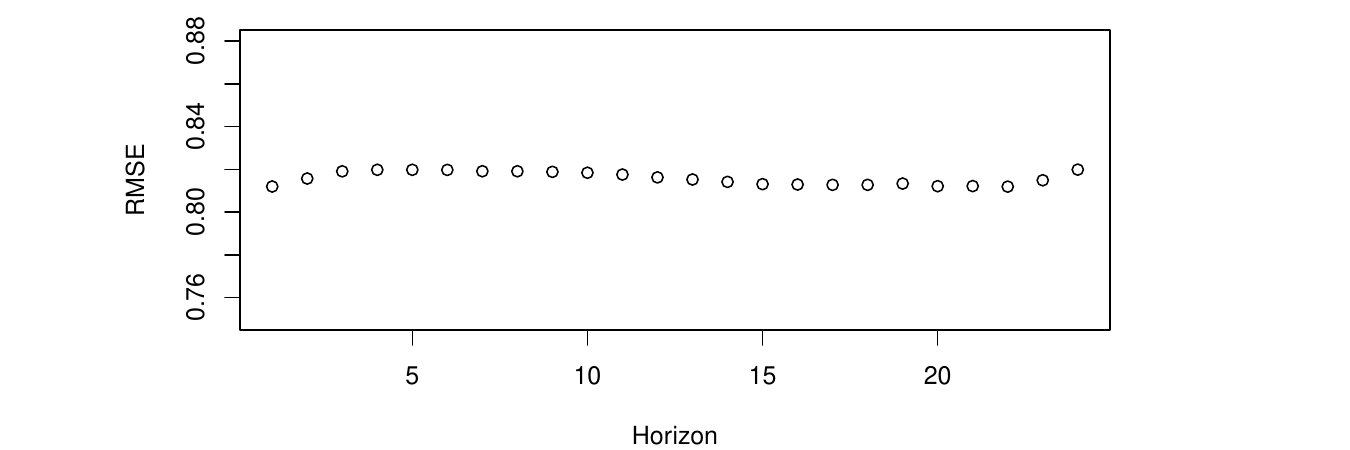} 
\end{knitrout}

\noindent Which is relatively constant. The offline parameters were optimized for $k=3$
and $k=18$, which can explain why it's not monotonic increasing with the horizon.


\section{Discussion and conclusion} \label{sec:summary}


\subsection{Extending functionality}
The current package is designed to make it easy to implement
new transformation functions and regression schemes, as well as using other
optimizers for tuning parameters.

Implementing a new transformation function is straight forward. It must receive
either a forecast matrix or a list of forecast matrices and return either after
processing. Furthermore, when used in an operational online setup, where the
transformation is executed when new data arrives, it's possible to save state
information in a transformation function, such that next time the function is
called, the state can be read and used. See the \code{lp()} function for
inspiration when writing a new transformation function.

A new regression scheme, e.g.\ a kernel or quantile regression, can be
implemented. A fitting function should be implemented in similar way as
\code{lm\_fit()} and \code{rls\_fit()}, such that the first argument is the
parameter vector and it returns a score value, which can be passed to an optimizer.

It's very easy to use other optimizers. The current fitting functions can simply
be passed to any optimizer in \Rprog, which follows the \code{optim()} way of
receiving a function for optimization, see the code in \code{lm\_optim()}.

In future versions new regression techniques, e.g.\ kernel regression (local
fitting) and quantile regression, might be added. The latter opens up the
possibilities to calculate probabilistic forecasts, see \citep{nielsen2006using}
and \citep{bjerregoard2021introduction}, as
well as carry out normalization and Copula transformations, which can be very
useful for spatio-temporal forecast models, see \citep{tastu2011spatio} or
\citep{lemos2021probabilistic}.

\subsection{Summary and conclusion}

This paper provides an entry point and reference for working with the
\onlineforecast package. The paper covers version 1.0 of the package, which has
been available on CRAN in almost one year at the time of writing.

The main contribution of the package is to make it easy to generate online
multi-step forecasts in a flexible way. The package contains functionalities not
directly available elsewhere, such as:
\begin{itemize}
\item Enabling the use of input variables given as forecasts, e.g.\ NWPs, in an
  easy and flexible way.
\item Modelling of dynamics and non-linearities using transformations including
  tuning the parameters of these transformations.
\item Recursive estimation for tracking time-varying systems computationally
  efficient for multiple horizons.
\end{itemize}
Furthermore, online operation with computationally effective updating
is uncomplicated -- so the package is well suited for real-time operational applications.

The \onlineforecast package has a significant value for anyone who needs to
model operational online forecasting. For example, in energy scheduling, where
recursive updated forecasts are needed as input to optimal decision making and
real-time control of systems. It can also be very useful for companies that need
online forecasts for other monitoring and real-time applications --
especially the functionality for model updating with very little computational
costs when new data becomes available, is a unique feature of the package.


\section*{Computational details}
We have tried to make the \onlineforecast package depend on as few other
packages as possible. Only a few additional packages are used in the core
functionalities: \pkg{R6} for the ``usual'' OOP functionalities and \pkg{Rcpp}
with \pkg{RcppArmadillo} for easy integration of fast compiled code. For
extending the modelling possibilities the \pkg{splines} and \pkg{pbs} packages
are essential, and for nice caching the \pkg{digest} package. We acknowledge the
\pkg{devtools} and \pkg{knitr}, \pkg{rmarkdown}, \pkg{R.rsp}, \pkg{testthat}
packages, which are indispensable for developing a package. We acknowledge
the \Rprog community and the amazing work behind \Rprog done by many people over the
years!

The results in this paper were obtained using \Rprog~4.1.3. \Rprog itself and all packages used are available from the
Comprehensive \Rprog Archive Network (CRAN) at \url{https://CRAN.R-project.org/}.

\section*{Acknowledgments}

The software has been developed with funding from multiple projects: FLEXCoop
(European Union’s Horizon 2020, grant agreement No 773909), Flexible Energy
Denmark, Heat 4.0 and Decision support tools for smart home energy management
systems (Innovation Fund Denmark, No. 9045-00017B, 8090-00046B and 8053-00156B),
TOP-UP (Innovation Fund Denmark and ERA-NET, No. 9045-00017B), Digital-twin, IEA
Annex 71 and 83 Danish participation (EUDP, No. 64019-0570, 64017-05139 and
64020-1007), and finally SCA+ (EU Interreg, No. 20293290).

\bibliography{onlineforecast}

\newpage

\begin{appendix}

\section{Forecast model notation}\label{sec:forec-model-notat}
In this section it is shown how to write \onlineforecast models in mathematical
notation. Both in a full description and how to write a shorter summarized
description. Note, that when variables are noted in bold font it indicates that
they are multi-variate.

A model can be described in full detail as presented in the following.

The transformation stage
\begin{alignat}{2}\label{eq:model-full-trans}
  \text{Intercept:}\quad& \mu_{t+k|t} &\,=\,& 1 \\[1ex]
  \text{Periodic:}\quad& \mB{x}\ns{per}{t+k|t} &\,=\,&  f\n{fs}(t;n\n{har}) \\[1ex]
  \text{Part 1:}\quad& x\ns{1}{t+k|t} &\,=\,& H(B;a) u\ns{1}{t+k|t} \\[1ex]
  \text{Part 2:}\quad& \mB{x}\ns{23}{t+k|t} &\,=\,&
  f\n{bs}(u\ns{2}{t+k|t};n\n{deg}) u\ns{3}{t+k|t} \\[1ex]
  \text{Part 3:}\quad& x\ns{4}{t+k|t} &\,=\,& u\ns{4}{t}
\end{alignat}
and the regression stage
\begin{align}\label{eq:model-full-reg}
  Y_{t+k|t} = \beta_{0,k} \mu_{t+k|t} + \mB{\beta}^T_{1,k}
  \mB{x}\ns{per}{t+k|t} + \beta_{2,k} x\ns{1}{t+k|t} + \mB{\beta}^T_{3,k}
  \mB{x}\ns{23}{t+k|t} + \beta_{4,k} x\ns{4}{t+k|t} + \varepsilon_{t+k|t}
\end{align}

Thus the model inputs are:
\begin{itemize}
\item $t$ is simply the time value.
\item $u\ns{1}{t+k|t}$ some forecast input (e.g.\ NWP variable).
\item $u\ns{2}{t+k|t}$ some forecast input (e.g.\ could be a deterministic
  value, e.g.\ time of day which is always know (the $|t$ could be omitted)).
\item $u\ns{3}{t+k|t}$ some forecast input (e.g.\ NWP variable).
\item $u\ns{4}{t}$ some value at time $t$ (e.g.\ an observation variable).
\end{itemize}
The functions which maps the inputs ($u$'s) to the regression inputs ($x$'s)
are:
\begin{itemize}
\item $f\n{fs}(t;n\n{har})$ is a function generating Fourier series of some
  implicit period length.
\item $H(B;a)$ is a low-pass filter.
\item $f\n{bs}(u\ns{2}{t+k|t};n\n{deg})$ is a function generating basis splines.
\end{itemize}
Their parameters are the transformation parameters:
\begin{itemize}
\item $n\n{har}$ is the number of harmonics.
\item $a$ is the low-pass filter coefficient.
\item $n\n{deg}$ is the degrees of freedom of the spline function.
\end{itemize}
which must be set or optimized.

The regression coefficients are
\begin{align}
  \mB{\beta}_k &= \left[\beta_{0,k} ~~ \mB{\beta}^T_{1,k} ~~ \beta_{2,k} ~~
    \mB{\beta}^T_{3,k} ~~ \beta_{4,k}\right]^T\\
  &=\left[ \beta_{0,k} ~~
    \beta_{1,1,k} ~~ \beta_{1,2,k} ~~ \dots ~~ \beta_{1,2n\n{har},k} ~~ 
    \beta_{2,k} ~~
    \beta_{3,1,k} ~~ \beta_{3,2,k} ~~ \dots ~~ \beta_{3,n\n{deg},k} ~~ 
    \beta_{4,k}\right]^T
\end{align}

If the model is fitted with a recursive scheme, thus the coefficients change
over time, it should be indicated by adding a $t$ to the subscript,
e.g.\ $\beta_{0,k,t}$. Furthermore, other parameters can exist, which can enter
an optimization at the transformation stage, e.g.\ the RLS forgetting factor
$\lambda$. The parameters which are optimized in the transformation stage
should be presented together.

\new{Specifying the model in all details can be cumbersome to include in some texts,
so it makes sense to simplify the notation. When using a simpler notation, as
suggested below, it should be stated, what is implicit (e.g.\ the regression
stage). Referencing the present text should be sufficient when using a simpler
notation. Naturally, all inputs, functions, etc.,\ should be described in some
way.

A model can be specified in a simpler way, e.g.\ the model above in one equation
\begin{align}
  Y_{t+k|t} =&~ \beta_{0,k} + \mB{\beta}^T_{1,k}\, f\ns{fs}{k}(t;n\n{har}) +
    \beta_{2,k}\, H_k(B;a) u\ns{1}{t+k|t} + \mB{\beta}^T_{3,k}\, f\ns{bs}{k}(u\ns{2}{t+k};n\n{deg}) u\ns{3}{t+k|t} \nonumber\\
  &+ \beta_{4,k}\, u\ns{4}{t} + \varepsilon_{t+k|t}
\end{align}
or writing the regression stage implicitly by removing the regression
coefficients where it's meaningful 
\begin{align}
  Y_{t+k|t} = \mu_k + f\ns{fs}{k}(t;n\n{har}) + H_k(B;a) u\ns{1}{t+k|t} + f\ns{bs}{k}(u\ns{2}{t+k|t};n\n{deg}) u\ns{3}{t+k|t} + \beta_k u\ns{4}{t} + \varepsilon_{t+k|t}
\end{align}
It's then implicit that the functions are different from the previous stated
functions, since they include the regression coefficients.} Again, if fitted with
a recursive scheme, then it can be indicated by adding a $t$ subscript,
e.g.\ $f\ns{fs}{k,t}(t;n\n{har})$.

To simplify further the $k$ on the functions can be implicit
\begin{align}
  Y_{t+k|t} = \mu + f\n{fs}(t;n\n{har}) + H(B;a) u\ns{1}{t+k|t} +
  f\n{bs}(u\ns{2}{t+k|t};n\n{deg}) u\ns{3}{t+k|t} + \beta u\ns{4}{t} + \varepsilon_{t+k|t}
\end{align}
and similarly the transformation parameters can be implicit
\begin{align}
  Y_{t+k|t} = \mu + f\n{fs}(t) + H(B) u\ns{1}{t+k|t} +
  f\n{bs}(u\ns{2}{t+k|t}) u\ns{3}{t+k|t} + \beta u\ns{4}{t} + \varepsilon_{t+k|t}
\end{align}
Then the functions and their parameters, and the fitting scheme (i.e.\ with
either LS or RLS for each horizon) should be described in some other way.

Finally, the most simplified notation would be to even remove the time indexing
\begin{align}
  Y = \mu + f\n{fs}(t) + H(B) u\n{1} +
  f\n{bs}(u\n{2}) u\n{3} + \beta u\n{4} + \varepsilon
\end{align}
after making clear how all the variables are defined.

\section{Transformation details} \label{app:transformation-of-inputs}

In this section a short introduction is given to the package functions for mapping
inputs in the transformation stage.

\subsection{Low-pass filtering}\label{sec:low-pass-filtering-appendix}

When modelling passive dynamical systems low-pass filtering is needed to
describe the input-output relation.  A simple first-order low-pass filter is
equivalent to the transfer function of a single resistor single capacitor
system -- equivalent to a first order ARX model.

The input time series is filtered with a transfer function
\begin{align}
  x_{t+k|t} = H(B;a) u_{t+k|t}
\end{align}
where the simplest first-order low-pass with stationary gain of one (unity DC gain) is
\begin{align}
  H(B;a) = \frac{1-a}{1-a B}
\end{align}
thus the low-pass filtered series becomes
\begin{align}\label{eq:first-order-lp}
  x_{t+k|t} = \frac{(1-a) u_{t+k|t}}{1-a x_{t-1+k|t-1}}
\end{align}
thus we have a filter coefficient $a$ between 0 and 1, which must be
tuned to match the particular time constant of the linear system. This filter is
implemented in the function \code{lp()}.

The following plot shows the response to a step function for different filter
coefficients $a$:
\vspace{-5ex}
\begin{center}  
\begin{knitrout}
\definecolor{shadecolor}{rgb}{0.969, 0.969, 0.969}\color{fgcolor}
\includegraphics[width=1\linewidth]{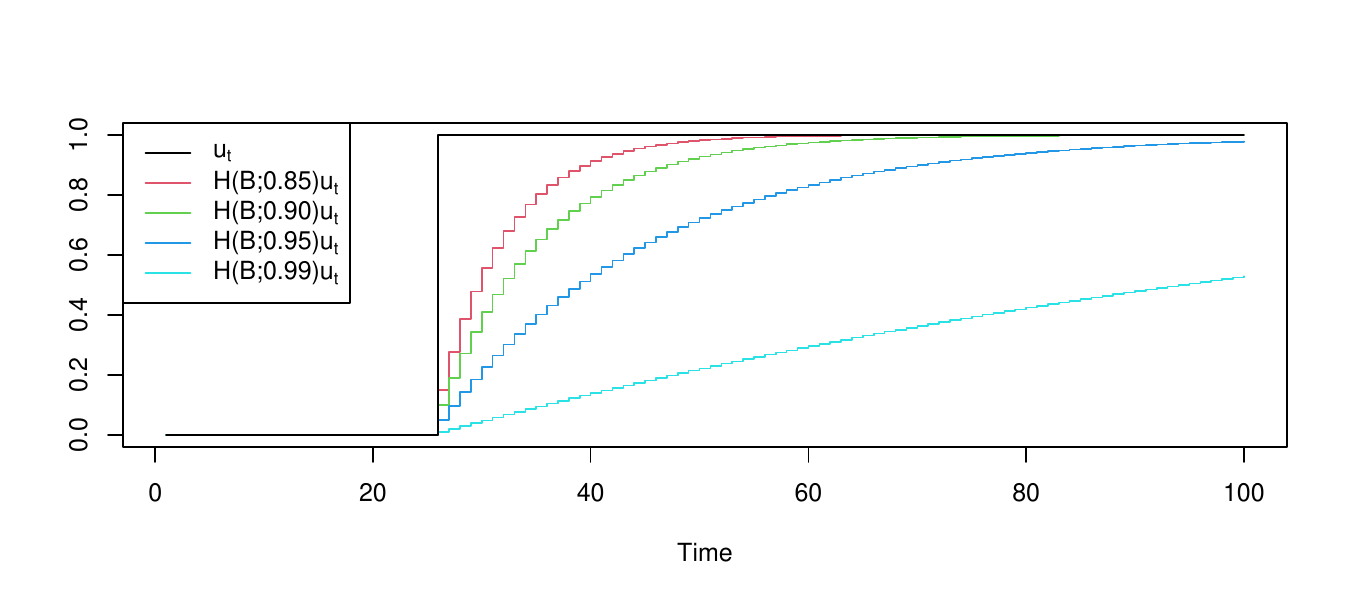} 
\end{knitrout}
\end{center}
\vspace{-5ex}
It is clearly seen that the responses are exponential functions with different
time constants depending on the value of $a$, such that a higher value results
in a slower response.


\subsection{Base-splines}\label{sec:base-splines-appendix}

A wide spread approach for modelling non-linear functional relations is to
apply spline basis functions \citep{hastie2009elements}. The basic idea is to
``resolve'' a single input into multiple inputs -- and use the them
instead of the single input in the regression. In this way a non-linear
function can be fitted between the single input and the model output.

The spline basis functions are calculated separately for each horizon
\begin{align}
  \mB{x}_{t+k|t} = f\n{bs}(u_{t+k|t}; n\n{deg})
\end{align}
where $n\n{deg}$ is the degree of the piecewise polynomial function, hence a
higher $n\n{deg}$ results in a more ``flexible'' function. Note, that $u_{t+k|t}$
is a single time series, but the $\mB{x}_{t+k|t}$ is a matrix of $n\n{deg}$
columns, i.e.\ as many times series all which will be used in the regression for
horizon $k$. The values where the piecewise polynomials meet, are known as
knots. The function to be used for transformations is \code{bspline()}, which
builds directly on the \code{bs()} function in \Rprog.

The input is resolved with spline basis functions, e.g.\ for an input
in the interval $[0,1]$ the basis splines for $n\n{deg}=4$ are:
\vspace{-5ex}
\begin{center}
\begin{knitrout}
\definecolor{shadecolor}{rgb}{0.969, 0.969, 0.969}\color{fgcolor}
\includegraphics[width=1\linewidth]{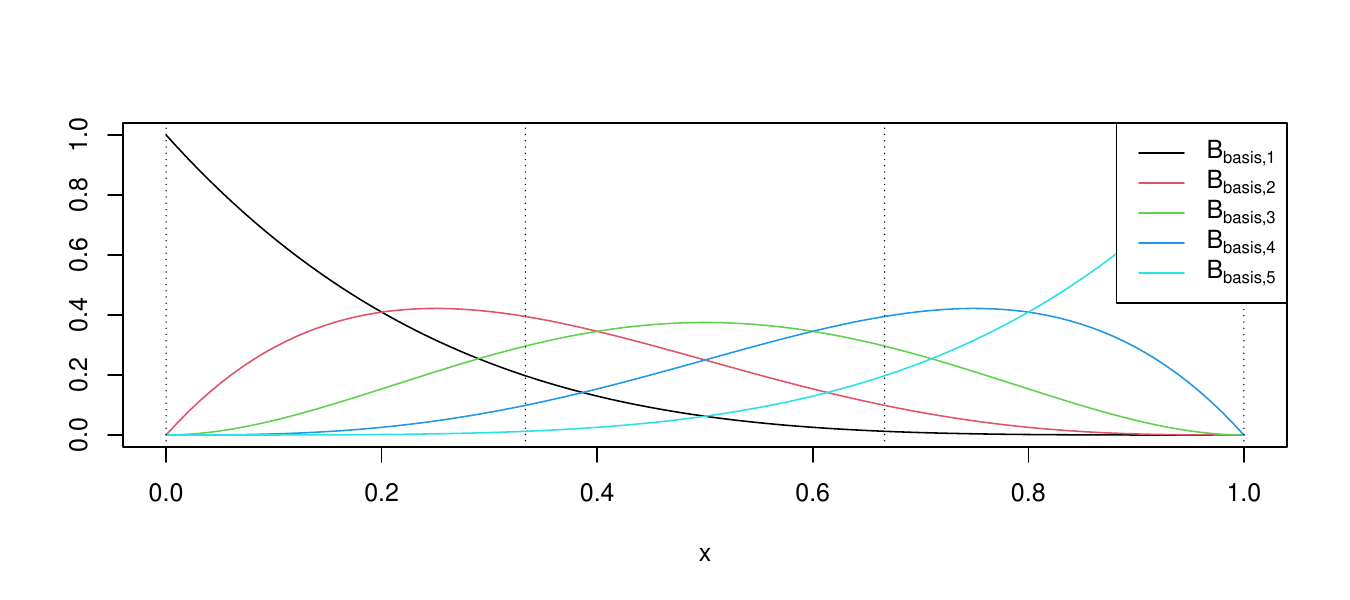} 
\end{knitrout}
\end{center}
\vspace{-5ex}
where the vertical dashed horizontal lines marks the knot points (must
be set in some way, usually set as equidistant quantiles of values of $u_t$).

An example of the resulting a spline functions, which can be fitted using
spline basis functions presented above is given. First, a non-linear function
with some added noise is simulated
\begin{align}
  Y_i = 
  \begin{cases} 
    u^3_i + \varepsilon_i \text{ for } u_i \leq 0\\
    u^3_i - 0.5 + \varepsilon_i \text{ for } u_i > 0
  \end{cases}
\end{align}
where $\varepsilon \sim N(0,0.1^2)$ and i.i.d.. A sample of 100
observations is simulated and spline basis functions of increasing
degree is generated and used as input to a linear regression
model. The resulting spline functions modelled is seen in the plot:

\begin{knitrout}
\definecolor{shadecolor}{rgb}{0.969, 0.969, 0.969}\color{fgcolor}
\includegraphics[width=1\linewidth]{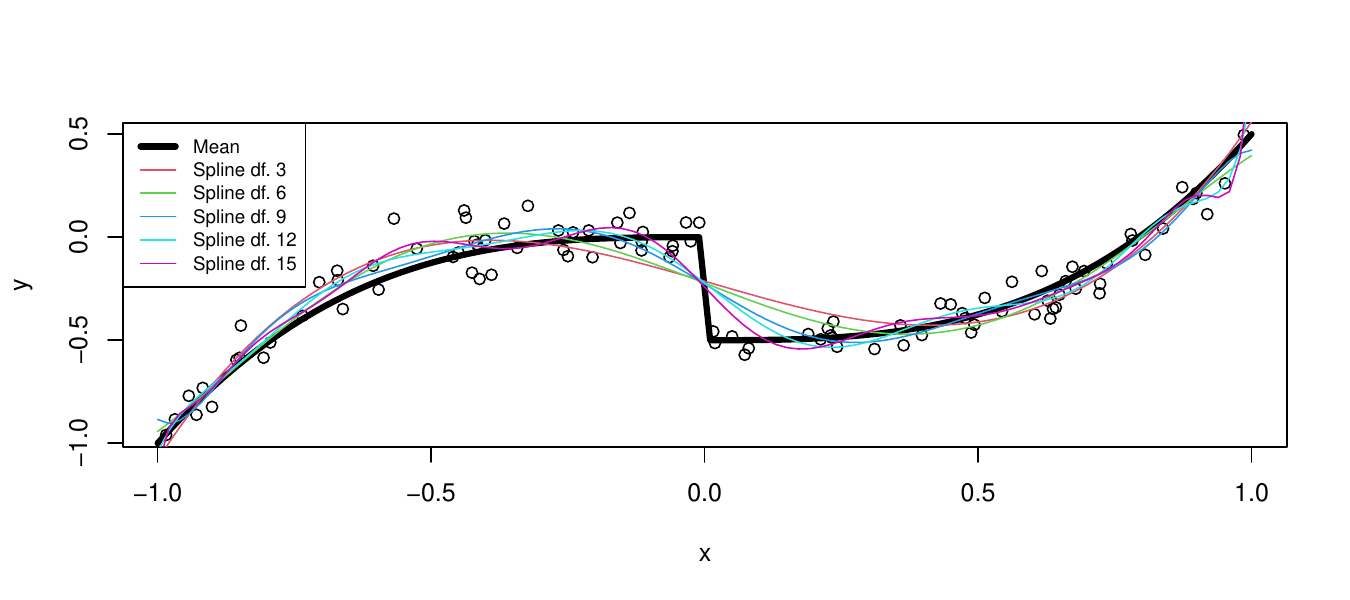} 
\end{knitrout}

It is clear that there is a balance between bias and variance:
A too low degree results in an under-fitted model (not able to
``bend'' enough), while a too high degree results in an over-fitted
model (bends to much). The degrees of freedom can be optimized using a
cross-validation approach.

\FloatBarrier

\subsection{Fourier series}\label{sec:fourier-series-appendix}

In order to model periodic phenomena a linear combination of Fourier
series is a very effective approach.

We use the notation
\begin{align}
  \mB{x}_{t+k|t} = f\n{fs}(u_{t+k|t}; n\n{har})
\end{align}
where $n\n{har}$ is the number of harmonic pairs included, hence
$\mB{x}_{t+k|t}$ is vector of length $2n\n{har}$, which is then linearly
combined in the regression stage. The package function for calculating Fourier series is \code{fs()}.

Exemplified with time as input $u_{t+k|t} = t+k$ (which is often done when
modelling a periodic phenomena, e.g.\ diurnal or yearly) the calculated series
is
\begin{align}
  \left[\sin\Big(\frac{2\pi}{t\n{per}} t\Big) ~~
  \cos\Big(\frac{2\pi}{t\n{per}} t\Big)
  ~~\sin\Big(\frac{2\cdot 2\pi}{t\n{per}} t\Big) ~~
  \cos\Big(\frac{2\cdot 2\pi}{t\n{per}} t\Big)
  ~~\dots~~
  ~~\sin\Big(\frac{n\n{har} 2\pi}{t\n{per}}
  t\Big) ~~ \cos\Big(\frac{n\n{har} 2\pi}{t\n{per}} t\Big) 
  \right]
\end{align}

An example of Fourier basis functions is plotted below for
$n\n{har}=3$, hence 3 pairs of harmonics and a period $t\n{per}=24$.

\begin{knitrout}
\definecolor{shadecolor}{rgb}{0.969, 0.969, 0.969}\color{fgcolor}
\includegraphics[width=1\linewidth]{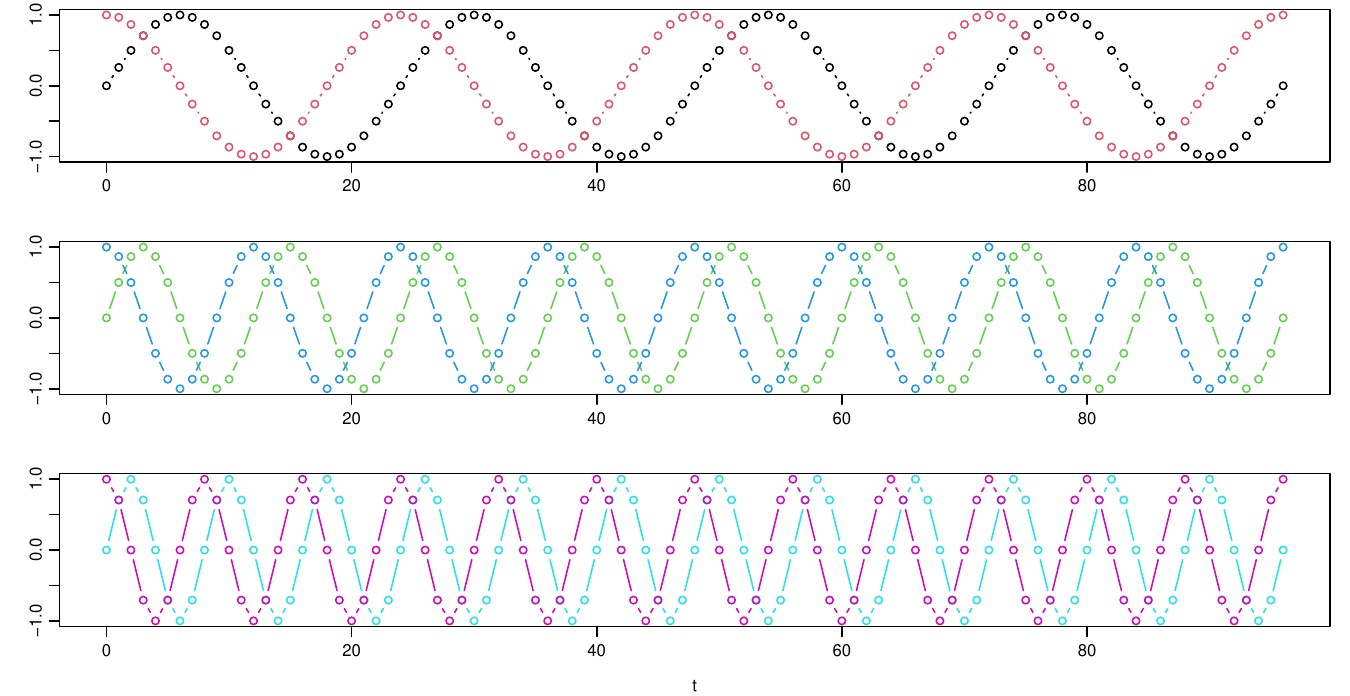} 
\end{knitrout}

\FloatBarrier

\subsection{Auto-regressive}
In classical time series analysis auto-regressive terms are included in models
to describe dynamical. This is equivalent to using low-pass filters as described
previously, however, it can often be useful to include an auto-regressive term
in a model, especially for forecasts a few steps ahead. In the package the
function \code{AR()} can be used to include the latest or lags of the current
observed model output. This can also be used for making an error model on
residuals.

\subsection{Nested transformations}
It is perfectly possible to combine transformations to create models involving
complicated functions. E.g.\ first resolve in basis splines and then low-pass
each of them
\begin{align}
  \mB{x}_{t+k|t} = H(B;a) f\n{bs}(u_{t+k|t}; n\n{deg})
\end{align}
This can be implemented in \Rprog by nesting the functions,
e.g.\ \code{lp(bspline())} (of course also providing the function arguments).

\subsubsection{\it Multiplication for coefficient-varying models}

In order to build coefficient-varying models, also called interaction effects
\citep{hastie1993varying}, it must be possible to multiply inputs. For the
forecast models it's carried out for each horizon separately
\begin{align}
  \mB{x}_{t+k|t} = u\ns{1}{t+k|t} \cdot u\ns{2}{t+k|t}
\end{align}
which usually in \Rprog can be carried out with the regular multiplication operator
'\code{*}'. However, an operator is needed for forecast matrices, which makes
sure that the correct horizons are multiplied and a bit more. The operator
'\code{\%**\%}' handles this and should be used multiplication in the  transformations.

\section{Regression} \label{app:regression}

In this section the two regression schemes implemented in \onlineforecast are
described. When fitting a model, thus estimating the regression coefficients,
data from a period $t \in (1,2,\dots,n)$ is used and passed on to either: the
\code{lm\_fit()} function which implements the Least Squares (LS) scheme, or the
\code{rls\_fit()} function, which implements the Recursive Least Squares (RLS) scheme.

One important difference between the two implementations is that in the LS the
coefficients are estimated using data from the entire period, thus they are
constant during the period and the calculated predictions are
``in-sample''. This is opposed to the RLS, where the coefficients are updated
through the period using only past data at each time $t$. In that case the
coefficients vary over time and the calculated predictions are
``out-of-sample''.

This difference is explained in the following and indicated by subscripting the
coefficient vector with $t$ only for the RLS.

\subsection{Least squares}

The regression coefficients for the $k$'th horizon is set in the vector
\begin{align}
  \mB{\beta}_{k} =\left[ \beta_{0,k} ~~
    \beta_{1,k} ~~ \dots ~~ \beta_{p,k} \right]^T
\end{align}
Note, that $t$ is not included in the subscript. 

The input data for the $k$ horizon is the design matrix
\begingroup
\setlength\arraycolsep{5pt} 
\renewcommand*{\arraystretch}{2} 
\begin{align}\label{eq:ls-designmatrix}
  \mB{X}_{k,t} = 
  \begin{bmatrix}
      x\s{0}{1+k|1} & x\s{1}{1+k|1} & \ldots &
      x\s{p}{1+k|1} \\    
      x\s{0}{2+k|2} & x\s{1}{2+k|2} & \ldots &
      x\s{p}{2+k|2} \\    
      \vdots & \vdots & & \vdots\\
      x\s{0}{n-1|n-1-k} & x\s{1}{n-1|n-1-k}  &
      \ldots & x\s{p}{n-1|n-1-k}\\
      x\s{0}{n|n-k} & x\s{1}{n|n-k} & \ldots & x\s{p}{n|n-k}\\
  \end{bmatrix}
\end{align}
\endgroup

The output observations are in the vector
\begin{align}
  \mB{y}_{k,n} =\left[ y_{1+k} ~~
    y_{2+k} ~~ \dots ~~ y_{n-1} ~~ y_{n}\right]^T
\end{align}

The LS estimates of the coefficients are
\begin{align}
  \hat{\mB{\beta}}_{k} = (\mB{X}_{k,n}\mB{X}_{k,n})^{-1} \mB{X}_{k,n}\mB{y}_{k,n}
\end{align}

The predictions are ``in-sample'' and calculated by
\begin{align}
  \hat{\mB{y}}_{k,n} = \mB{X}_{k,n} \hat{\mB{\beta}}_{k}
\end{align}
and returned when fitting a model with \code{lm\_fit()}.

The estimated coefficients may now be used for ``out-of-sample'' prediction (for
$t\n{new} \geq n$), with the input
\begin{align}
  \mB{x}_{t\n{new}+k|t\n{new}} = \left[ x\s{0}{t\n{new}+k|t\n{new}} ~~ x\s{1}{t\n{new}+k|t\n{new}} ~~ \dots ~~
    x\s{p}{t\n{new}+k|t\n{new}} \right]^T
\end{align}
by
\begin{align}
  \hat{y}_{t\n{new}+k|t\n{new}} = \mB{x}_{t\n{new}+k|t\n{new}} \hat{\mB{\beta}}_{k} 
\end{align}
This can be done by providing new data to the \code{lm\_predict()} function.

\subsection{Recursive least squares}

In the RLS scheme the coefficients are recursively updated through the
period. Time $t$ steps from 1 to $n$ and in each step the ``newly'' obtained
data at $t$ is used for calculating updated coefficients. The coefficient vector
has the same structure as for LS
\begin{align}
  \mB{\beta}_{k,t} =\left[ \beta_{0,k,t} ~~ \beta_{1,k,t} ~~ \dots ~~ \beta_{p,k,t} \right]^T
\end{align}
The only difference is that we now subscript with $t$ because it varies over
time.

Only the most recent input data at $t$ (the row at $t$ from the LS design
matrix in Equation \eqref{eq:ls-designmatrix}) is used in each update
\begin{align}
  \mB{x}_{k,t} = \left[ x\s{0}{t|t-k} ~~ x\s{1}{t|t-k} ~~ \dots ~~
    x\s{p}{t|t-k} \right]^T
\end{align}  
and similarly the most recent output observation $y_{t}$. At each time $t$ the coefficients are updated by
\begin{align}
  \mB{R}_{k,t} &= \lambda \mB{R}_{k,t-1} + \mB{x}_{k,t} \mB{x}^T_{k,t}\\
  \hat{\mB{\beta}}_{k,t} &= \hat{\mB{\beta}}_{k,t-1} + \mB{R}^{-1}_{k,t}
  \mB{x}_{k,t} (y_t - \mB{x}^T_{k,t} \hat{\mB{\beta}}_{k,t-1})
\end{align}

Hence, when applying RLS for data from the period $t \in (1,2,\dots,n)$ the
RLS provides a new value of the coefficients for each time $t$ (opposed to LS).

The predictions are calculated recursively as well by using the updated
coefficients at each time $t$. Given the inputs
\begin{align}
  \mB{x}_{t+k|t} = \left[ x\s{0}{t+k|t} ~~ x\s{1}{t+k|t} ~~ \dots ~~
    x\s{p}{t+k|t} \right]^T
\end{align}
the prediction is
\begin{align}
  \hat{y}_{t+k|t} = \mB{x}_{t+k|t} \hat{\mB{\beta}}_{k,t}
\end{align}
Only past data has been used when calculating the predictions through the
period, hence they are ``out-of-sample'' predictions (these predictions are returned by \code{rls\_fit()}).

The initial value of $R$ is set simply set to a zero matrix with diagonal
$1/10000$ and $\beta$ set to a zero vector.

An alternative updating scheme, which is actually the implemented scheme (gives
the same results as the scheme above), is the Kalman gain scheme
\citep{sayed1994state}, where matrix inversion is avoided
\begin{align}
  \mB{K}_{k,t} &= \frac{\mB{P}_{k,t-1} \mB{x}_{k,t}}{ \lambda + \mB{x}^T_{k,t}
    \mB{P}_{k,t-1} \mB{x}_{k,t}}\\
  \hat{\mB{\beta}}_{k,t} &= \hat{\mB{\beta}}_{k,t-1} + \mB{K}_{k,t} (y_t -
  \mB{x}^T_{k,t} \hat{\mB{\beta}}_{k,t-1})\\
  \mB{P}_{k,t} &= \frac{1}{\lambda} \left( \mB{P}_{k,t-1} - \mB{K}_{k,t}
  \mB{x}^T_{k,t} \mB{P}_{k,t-1} \right)
\end{align}
This actually opens up the possibilities for self-tuned variable forgetting
\citep{shah1991recursive}.

\end{appendix}

\address{
Peder Bacher\\
Dynamical Systems\\
Department of Applied Mathematics and Computer Science\\
Technical University of Denmark\\
Asmussens Allé, Building 303B\\
2800 Kgs. Lyngby, Denmark\\
E-mail: \email{pbac@dtu.dk}\\
URL: \url{https://www.compute.dtu.dk/english/research/research-sections/dynsys/}
}

\address{
Hjörleifur G. Bergsteinsson\\
Dynamical Systems\\
Department of Applied Mathematics and Computer Science\\
Technical University of Denmark\\
Asmussens Allé, Building 303B\\
2800 Kgs. Lyngby, Denmark\\
E-mail: \email{hgbe@dtu.dk}\\
URL: \url{https://www.compute.dtu.dk/english/research/research-sections/dynsys/}
}

\address{
Linde Frölke\\
Dynamical Systems\\
Department of Applied Mathematics and Computer Science\\
Technical University of Denmark\\
Asmussens Allé, Building 303B\\
2800 Kgs. Lyngby, Denmark\\
E-mail: \email{hgbe@dtu.dk}\\
URL: \url{https://www.compute.dtu.dk/english/research/research-sections/dynsys/}
}

\address{
Mikkel L. Sørensen\\
Dynamical Systems\\
Department of Applied Mathematics and Computer Science\\
Technical University of Denmark\\
Asmussens Allé, Building 303B\\
2800 Kgs. Lyngby, Denmark\\
E-mail: \email{mliso@dtu.dk}\\
URL: \url{https://www.compute.dtu.dk/english/research/research-sections/dynsys/}
}

\address{
Julian Lemos-Vinasco\\
Dynamical Systems\\
Department of Applied Mathematics and Computer Science\\
Technical University of Denmark\\
Asmussens Allé, Building 303B\\
2800 Kgs. Lyngby, Denmark\\
E-mail: \email{jlvi@dtu.dk}\\
URL: \url{https://www.compute.dtu.dk/english/research/research-sections/dynsys/}
}

\address{
Jon Liisberg\\
Dynamical Systems\\
Department of Applied Mathematics and Computer Science\\
Technical University of Denmark\\
Asmussens Allé, Building 303B\\
2800 Kgs. Lyngby, Denmark\\
E-mail: \email{jlvi@dtu.dk}\\
URL: \url{https://www.compute.dtu.dk/english/research/research-sections/dynsys/}
}

\address{
Jan Kloppenborg Møller\\
Dynamical Systems\\
Department of Applied Mathematics and Computer Science\\
Technical University of Denmark\\
Asmussens Allé, Building 303B\\
2800 Kgs. Lyngby, Denmark\\
E-mail: \email{jkmo@dtu.dk}\\
URL: \url{https://www.compute.dtu.dk/english/research/research-sections/dynsys/}
}

\address{
Henrik Aalborg Nielsen \\
ENFOR A/S\\
Røjelskær 11, 3. \\
2840 Holte, Denmark\\
E-mail: \email{han@enfor.dk}\\
URL: \url{https://www.enfor.dk}
}

\address{
Henrik Madsen\\
Dynamical Systems\\
Department of Applied Mathematics and Computer Science\\
Technical University of Denmark\\
Asmussens Allé, Building 303B\\
2800 Kgs. Lyngby, Denmark\\
E-mail: \email{hmad@dtu.dk}\\
URL: \url{https://www.compute.dtu.dk/english/research/research-sections/dynsys/}
}

\end{article}

\end{article}

\end{document}